\documentclass{aa}
\usepackage{graphics}
\begin{document}
\thesaurus{03.09.4;03.13.5;03.19.3}

\title{Simulations of Observations with the Optical Monitor of the X-Ray Multi-Mirror Satellite}


\author{P.\ Royer \thanks{Aspirant au Fonds National de la 
Recherche Scientifique (Belgium)} 
\and J.\ Manfroid \thanks{Directeur de Recherche au Fonds National de la 
Recherche Scientifique (Belgium)} 
\and E.\ Gosset \thanks{Chercheur Qualifi\'e au Fonds National de la 
Recherche Scientifique (Belgium)} 
\and J.-M.\ Vreux
} 

\offprints{P. Royer, proyer@ulg.ac.be}

\institute{Institut d'Astrophysique, Universit\'e de Li\`ege,
5, Avenue de Cointe, B-4000 Li\`ege, Belgium}
\date{Received date; accepted date}
\maketitle

\markboth{Properties of the XMM-OM photometry}{Properties of the XMM-OM photometry}

\newcommand{\ico}{$ic_{\rm 0}$}
\newcommand{\icr}{$ic_{\rm red}$}

\begin{abstract}
This paper addresses the question of the observations
to be performed with the Optical Monitor (OM) of the 
X-ray Multi-Mirror Satellite (XMM) under several aspects. 
First, we discuss XMM-OM's photometric system and its 
colour transformations towards the standard $U\!BV$ system. 
Second, we establish a set of procedures to determine 
the temperature and the amount of interstellar absorption 
affecting the observed stars. Last, we address the possibility 
of isolating quasars in multidimensional colour diagrams 
based on the XMM-OM filter set.

\keywords{Satellite : XMM -- Photometry : colour transformations; 
Interstellar absorption; Quasars : identification of candidates}
\end{abstract}

%
%
\section{Introduction}

Thanks to its unprecedented large X-ray collecting power, 
the XMM satellite (for X-ray Multi-Mirror) is expected to 
discover a wealth of new X-ray sources. 
In order to allow a quick and reliable identification of the sources, 
as well as to enable multiwavelength monitoring of their variability, 
the satellite is equipped with an optical complement known as the optical
monitor, XMM-OM. This particular element is co-aligned with the three
X-ray telescopes and consists of a 30~cm Ritchey-Chr\'etien telescope 
coupled with a photon-counting detector. The latter
consists of a photocathode followed by three micro-channel
plates as intensifier and by a tapered fiber-optic bundle connected to a
fast scanning CCD. The usable area is made of 256 by 256 physical pixels
but a centroiding process locates the events to 1/8 of a pixel
thus mimicking a 2048 by 2048 device. 
The field of view is 17 arcmin by 17 arcmin 
with a centroided pixel size of 0.5 arcsec. 
At the end of the OM exposure, the cumulated image is downloaded to
the ground. An engineering mode allows to transmit the whole
image but the routine science mode necessitates predefined windowing 
and/or binning.
The telescope was designed so that the limiting magnitude would be 
no less than 24 when working in unfiltered light. More details can be found in Fordham et al. 1992 and Mason et al. 1996.

XMM-OM holds a 6-filter, UV and optical, 
photometric system. Three bands of the system were designed 
to match the Johnson's $U\!BV$ system (Johnson 1955; Bessel 1990).
The rectangular profiles of the latter filters render the colour 
transformations between the XMM-OM and the Johnson systems 
quite complicated (see Royer \& Manfroid 1996 for a discussion 
on the untransformability of rectangularly shaped filters), 
or even not always possible. Such colour transformations will 
nevertheless remain necessary for those who will have to 
compare XMM-OM observations with ground-based ones. 
In this paper, we provide the reader with theoretical estimations 
of these colour transformations.

The XMM-OM filter set comprises three non-standard filters, exploring
a wavelength  domain  (UV) where no  extensive observations  have been
performed  to date.  Combined with the  poor  match between the XMM-OM
optical filters and  the  classical ground-based  standard photometric
systems, this fact enhances the  importance of testing our ability  to
determine some physical properties  of the observed stars directly  in
the natural XMM-OM colour system.  In the present  paper, we will show
that it is possible to estimate both the temperature and the amount of
interstellar absorption (or ``reddening'') for the hot stars that will
be observed with XMM-OM.

The  natural  XMM-OM  colour  system   can  also  be  used  for  other
investigations.  For example,  we have  explored  the possibilities to
discriminate   quasars from stars  in  multidimensional  colour spaces
based on the XMM-OM photometry.  Nevertheless, the goal of this  paper
is not so  much to provide  the reader with exact analytical relations
for these matters as to give qualitative  results that will tell him
how to observe and  treat the data in order  to get the best outcome,
and what kind of results are to be expected.

\begin{figure*}[tbh]
\resizebox{\hsize}{!}{\rotatebox{270}{\includegraphics{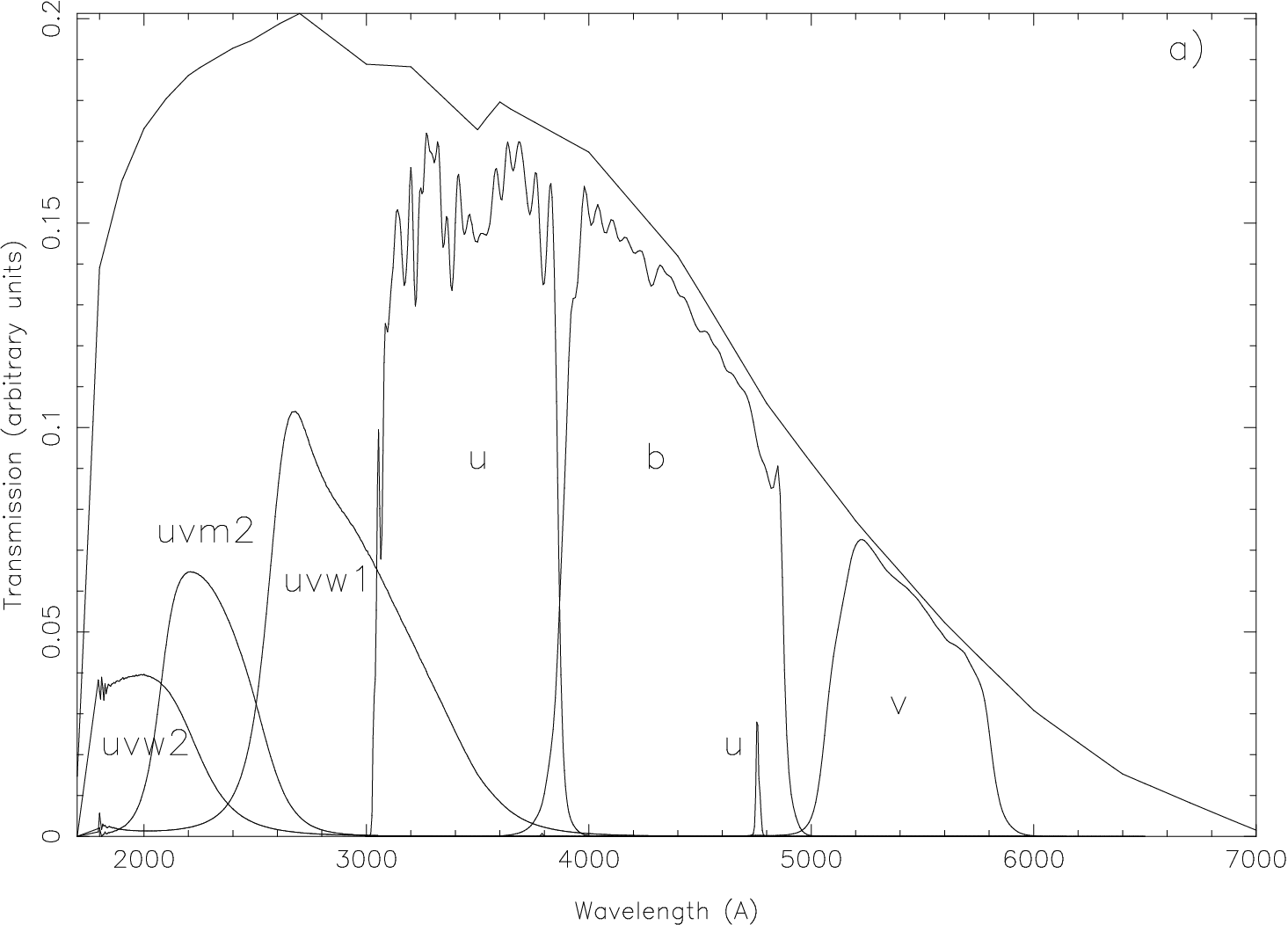}}\rotatebox{270}{\includegraphics{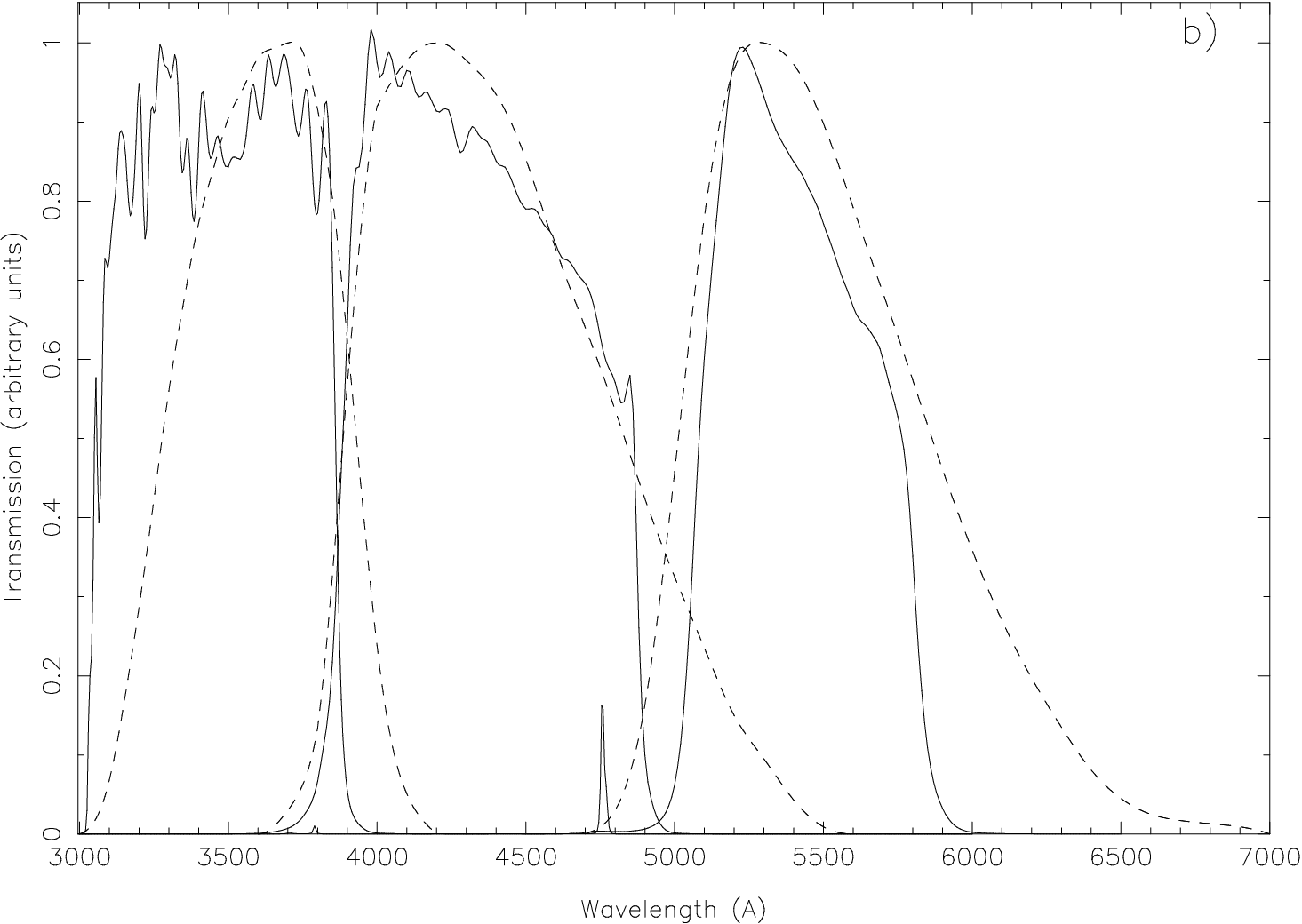}}}
\caption{{\bf{a}} XMM-OM's photometric system. The upper curve is the 
instrumental response in the absence of filter. It results from the product 
of the response curves of the detector, of the three mirrors and 
of the detector window. {\bf{b}} Comparison of the XMM-OM optical 
and the Johnson $U$, $B$ and $V$ filters. The pairs of filters are, 
from left to right: $u$ and $U$; $b$ and $B$; $v$ and $V$. 
The dashed curves are the Johnson filters.}
\end{figure*}

In Sect.~2, we present the tools we used for the synthetic photometry.
In Sect.~3, we discuss the colour transformations between 
the XMM-OM optical filters and the Johnson $U$, $B$, $V$ system. 
We treat the temperature and the interstellar absorption determination 
in Sects.~4 and 5 whereas the selection of quasar candidates is addressed 
in Sect.~6. Sect.~7 outlines the main conclusions of our study.

\section{The basic information and hypotheses}
\subsection{The filters}
All simulations presented hereafter were 
established on the basis of synthetic photometry. The filter 
passbands that we used for the XMM-OM photometric system are 
extracted from the XMM Users' Handbook (Dahlem \& Schartel 1999). 
Their transmission curves result from the product of the 
transmission curves and/or sensitivity curves of all physical 
devices in the XMM-OM light path. Namely, we have the filters themselves, 
three aluminized mirrors, one window at the entry of the 
detector and the detector response curve. All these curves originate 
from the same reference as above. The resulting passbands are plotted 
in Fig.~1a and their physical characteristics are summarized in Table~1. Throughout the rest of this paper, small characters refer to 
the XMM-OM filters and capital letters refer to the standard ones 
(i.e. the $U$, $B$ and $V$ taken from Bessel 1990).

\begin{table}[hbt]
\begin{center}
\begin{tabular}{|c|c|c|c|}
\hline
     filter   &     $\lambda_{o}$ &     $\lambda_{\rm max}$ &     FWHM \\
\hline
 {\it{uvw2}}   & 2070  & 2000 & $\sim\,500$ \\
\hline
 {\it{uvm2}}   & 2298  & 2210 & 439 \\
\hline
 {\it{uvw1}}   & 2905  & 2680 & 620 \\
\hline
 {\it{u}}      & 3472  & 3270 & 810 \\
\hline
 {\it{b}}      & 4334  & 3980 & 976 \\
\hline
 {\it{v}}      & 5407  & 5230 & 684 \\
\hline
\end{tabular}
\caption{Main characteristics of the XMM-OM filters. The wavelengths
are expressed in \AA. $\lambda_0$ is the effective wavelength and $\lambda_{\rm max}$ is the position of the maximal transmission.}
\end{center}
\end{table}

The XMM-OM filter set consists of 3 UV and 3 optical filters, 
further denoted by $uvw2$, $uvm2$, $uvw1$ and $u$, $b$, $v$ respectively, 
in order of increasing central wavelength. The $u$, $b$ and $v$ filters are
intended to match the Johnson $U$, $B$ and $V$ (Bessel 1990) filters 
respectively, but they possess more or less rectangularly 
shaped passbands, which greatly affects the quality 
of the match and of the subsequent 
colour transformations between the two systems. 
This is especially true for what concerns the (mis)match between the 
Johnson $U$ and XMM-OM $u$ filters and thus the $(U-u)$ and $(U-B)$ related 
colour transformations. The XMM-OM $u$ band is characterized by:
\begin{itemize}
\item a strong sensitivity in the 3100-3300~\AA\ region, where the ground-based Johnson's $U$ is essentially blind;
\item a  reduced sensitivity in the  critical Balmer  decrement region
(3850-4100~\AA),  leading to untransformability between both bands
for stars  showing Balmer jump (see Sects.~3.1.1, 3.1.6, 3.1.7).  As we show
later, $(U-u)$ can be evaluated thanks to $uvw1$.
\end{itemize}

Fig.~1b allows comparison between XMM-OM and standard $U$, $B$ and $V$ filters.

The UV XMM-OM  filters do not correspond  to any pre-existing standard
system. No transmission measurements below 1800 \AA\  are given in the
XMM  flight  simulator software  files,   so  we added  a  point  with
$\sim0\%$ transmission at 1700 \AA\ in the UV filters.

\subsection{The stellar spectra}
The spectra  we used  for the  synthetic photometry are  those  of the
Kurucz' ATLAS 9  atmosphere  models (Kurucz   1992),  reddened by  the
Cardelli interstellar extinction model (Cardelli et   al.  1989)  with  an average
reddening   law   ($R_{\rm V}~=~A_{\rm   V}/E(B-V)~=~3.1$)  when
necessary.  Except when otherwise stated, the set of spectra we chose
mimics an unreddened solar composition main  sequence (MS) since these 55
spectra possess    the following
characteristics:   $A_{\rm V}~=~0.$;  $[\rm Fe/H]~=~0.$; $\log~g~=~4.$;
$3500\,$K$~\leq~T_{\rm eff}~\leq~35000\,$K.  Stars we refer to as {\it
giants} have  $\log~g~=~3.5$ (Allen, 1976),  those we refer to as {\it
metal poor stars} have  $[\rm Fe/H]~=~-1.5$, typical for  the halo of  our
Galaxy.   The temperature interval between  two adjacent Kurucz models
varies  along  the tracks:  it  is $250~$K  below  $10000$~K,  $500$~K
between  $10000$~K and  $13000$~K,   $1000$~K  between  $13000$~K  and
$35000$~K, and $2500$~K between $35000$~K and $50000$~K.

Throughout the rest of this paper, the MS spectrum with 
$T_{\rm eff}~=~9500\,$K was attributed a magnitude of zero in all filters.

All colour transformations established on the basis of the Kurucz
spectra were compared with their equivalent obtained with the stellar
spectral atlas prepared by Fioc \& Rocca-Volmerange (1997, hereafter
FRV). This atlas comprises 65 spectra of stars of various luminosity
classes (3 supergiants, 30 giants and 32 dwarfs) and spectral types
($T_{\rm eff}~\in~[2500,180000]$~K). It was constructed from observed
spectra whenever possible (Gunn \& Strycker 1983; Heck et al. 1984),
from synthetic spectra otherwise (Kurucz  1992; Clegg \& Middlemass
1987).  

\subsection{The stellar locus for the quasar fields}
In Sect.~6, we analyse the ability of the XMM-OM photometric system to
discriminate between quasars and stars.  Therefore, we have to delimit
both   the quasar   (see Sect.~2.4)  and    the stellar  loci in   the
multidimensional  colour  space defined by  the   XMM-OM system.   The
stellar locus   is  defined by   all  the   stars that  are  potential
contaminants of the quasar candidate population.  When looking at high
galactic latitudes,  the field stars are halo  main sequence stars. We
modelize their colours by integrating the  emergent fluxes from Kurucz
model atmospheres.  We represent halo main sequence stars by the whole
range of $T_{\rm   eff}$,  $\log~g~=~4$ and $[\rm Fe/H]~=~-1.5$.   Due  to
evolution,   only   the part   for   which $T_{\rm  eff}~\leq~7000\,$K
(corresponding to an early F  spectral type) is still populated. These
latter stars  are  the    major  constituents of  the   field  stellar
population.  The reason  why we consider  the hotter  stars as well is
that their colours  are close to those of  other families of stars. We
will use that property in Sect.~6, and discuss it further there.

Disk main sequence stars are also a  possible contaminant in a list of
quasar candidates.    However,     only  the    cool   end    ($T_{\rm
eff}~\leq~7000\,$K) could be a problem and is  taken into account.  We
chose  to represent  them by  a $\log~g~=~4$ and  a solar metallicity.
Indeed,  the disk is rather  thin: it has a typical scale height of 1 kpc,
which corresponds to a distance  modulus of 10 magnitudes.  This means
that disk   OBA stars are   too  bright in  apparent magnitudes  to be
mistaken for quasars.  Another possible  contaminant is constituted by
halo giants.  We  represent them by Kurucz  models  with the following
parameters:    $T_{\rm  eff}~\leq~7000\,    $K, $\log~g~=~3$       and
$[\rm Fe/H]~=~-1.5$.

Some hotter stars are also present at high galactic latitudes as trace
constituents  but  since they  are  bluer, a  property they share with
low-redshift quasars, they could constitute  a strong contaminant in a
quasar candidate  list. We consider three families.   The first one is
the Horizontal-Branch   (HB)  BA  stars.  These stars     have $T_{\rm
eff}~\geq~10000\, $K (up to $\sim  30000\,$K).  On the HB, the gravity
is strongly dependent on  the  effective  temperature:  we chose  the
dependency law  we derived by averaging  the ones quoted by Moehler et
al.\ (1999) and  by Conlon et  al.\ (1991).   According  to Moehler et
al.\ (1999), the  best fit of Kurucz  models  to HB star's spectra  is
obtained   for  metal  rich  chemical  compositions. We  adopted $[\rm
Fe/H]~=~0.5$.  As noticed by Miller \& Mitchell (1988), the population
of HB stars  corresponds  to   intrinsically  bright objects and    is
observed to tail-off at faint  magnitudes due to the finite  dimension
of the halo.

A  second family is constituted of  the subdwarfs  of spectral type OB
(sd  OB).  Their evolutionary status  is  still somewhat uncertain but
they are  usually associated to the  Extended Horizontal Branch and to
the  evolution thereof (see Caloi,  1989), although some authors refer
to  some of them  as  being post-AGB objects.    We represent  them by
Kurucz models with   $T_{\rm eff}$ between  $20000\,$K and  $50000\,$K
(see Conlon  et  al.\ 1991 but  also  Table 1 of  Lenz  et al.\ 1998),
$\log~g~=~5$ and solar metallicity.

Finally, the third  family is made of  degenerate stars, the so-called
white dwarfs. Some  models of  degenerate stars exist  but only  a few
have  their emergent  flux    published. As  a  first   approximation,
degenerate stars  are known  to have $U\!BV$   colours very similar to
black bodies.  We therefore computed the colours  of black bodies.  In
any case, this  characteristic is  perhaps not  general and  does  not
apply to the  UV part of  the  spectrum. We finally  used the emergent
fluxes of the pure-hydrogen atmosphere  models for degenerate stars of
Koester (1999).   We restricted  ourselves  to  effective temperatures
ranging  from $7000$~K  to $80000$~K  and   to $\log~g~=~8.5$.  We  also
integrated  the   spectra      of the    four   white-dwarf    primary
spectrophotometric standards described by   Bohlin et al.\ (1995)  and
the  models  of  Wesemael  et  al.\ (1980).  Both   works are  in good
agreement  (although   not  necessarily  independent)  with  Koester's
models.

\subsection{The quasar spectrum}
The  quasar  spectra used   in  Sect.~6 were derived   from an average
spectrum build from  the composite spectra of  Zheng et al. (1997) and
Francis et al. (1991).  The former was used from  310 to 2000~\AA\ and
the latter from  2000 to 6000~\AA. The match  between  both spectra is
reasonably good, as  shown by Zheng et  al.  (1997, their Fig.~9).  It
was  performed  on the  continuum windows identified  by these authors
between  1400 and 2200~\AA.    The absorption characteristics of   the
quasar  spectra    due  to  the intervening    Ly$\alpha$  clouds were
established, between Ly$\alpha$\ and the Lyman  break, on the basis of
the concept  of Oke \& Korycansky (1982)  and of the works presented by
Irwin et al.  (1991), Zuo \&  Lu (1993) and  Warren et al. (1994).  For
the  region below  the   Lyman  break,  three absorption models   were
computed. The first two   were adapted from  the works of M\o ller  \&
Jakobsen (1990), M\o ller \& Warren  (1991), Warren et al.  (1994) and
Giallongo \& Trevese (1990).   The first one  only takes into  account
the  hydrogen   clouds with column    densities inferior  or  equal to
$10^{17}~{\rm  cm}^{-2}$   (sometimes called  the   Ly$\alpha$  forest
clouds).  It will further  be referred to  as model~A. The second  one
additionally includes systems with  column densities  between $10^{17}$
and $10^{20}~{\rm cm}^{-2}$      (sometimes called the  Lyman    limit
systems).  It will further be  referred to as model  B.  As an extreme
case, we also  considered a  model where  a strong (10 magnitudes) absorption
occurs in the Lyman continuum  at a redshift  very close to the one of
the quasar,  absorption which persists in   the whole observable Lyman
continuum.  This model will further be referred  to as model~C.  These
three models are strongly inspired by those  described by Royer (1994)
where full details can be found.   To fix ideas,  it is worth noticing
that the  higher normalization of Madau (1995)  puts  his model between
our models B and C.

It  must  be clear   that the  quasar  spectrum  defined above is only
representative of an average  quasar  and that some individual  quasar
spectra deviate strongly from it.  The true population of quasars will
exhibit some dispersion around the characteristics of this quasar.  It
is well known (see e.g.\ Francis et  al.\ 1991) that the population of
quasars displays a variety of power-law flux distributions and that the
emission-line equivalent widths vary from one object to the other both
in a systematic way  (the Baldwin effect) and  in a random way. On the
basis of the power-law index  dispersion reported, e.g.\ by Francis et
al.  (1991), we expect this effect to spread the colours of quasars by
$\pm~0.3$ mag   around  our  tracks.   In addition, particular
realizations of the distribution  of high column density  clouds along
some line of sight could induce strong deviations from the mean 
behaviour.

\section{Colour transformations}
Colour transformations between a physical and a standard filter set
always are a source of photometric errors (see  e.g. Sterken \&
Manfroid 1992 or Royer \& Manfroid 1996 and references therein). In
the  particular case of the XMM-OM optical filters, the important
discrepancies between the standard and non-standard $U$, $B$ and $V$ filters
renders the choice of adequate transformations even more critical. In
order to avoid additional errors introduced by subsequent colour computations, we established colour transformations for both single
filters and colour indices.

\begin{table}[hbt]
\begin{center}
\begin{tabular}{|c|c|c|c|c|c|c|}
\hline
           & \it U--u & \it B--b & \it V--v & \it U--B & \it B--V & \it U--V\\
\hline
\it u--b    &  202    &  102    &         &  102    &         &         \\
\hline
\it uvw1--u &  222    &         &         &         &         &         \\
\hline
\it b--v    &         & 211 / 3 & 111 / 2 &         &    1    &         \\
\hline
\it uvw1--b &         &         &         &  102    &         &         \\
\hline
\it u--v    &         &         &         &         &         & 101 / 1 \\
\hline
\it uvw1--v &         &         &         &         &         & 102 / 1 \\
\hline
\end{tabular}
\caption{Summary of the colour transformations established in this
paper. Capital letters refer to Johnson  $U$, $B$ and $V$ filters. 
A one, two or three digit code {$i_1,...,i_n$} 
characterizes the transformation: $n$ is the number of 
temperature domains that have to be considered; $i_k$ 
is the order of the transformation in the $k^{\rm th}$ 
domain (0 means no transformation is possible), e.g. 102 
indicates that one has to divide the temperature domain in 3
distinct parts in which respectively linear, no and second order 
transformations are possible. The numbering of the domains is 
going from the hottest to the coolest stars.}
\end{center}
\end{table}

We  restricted ourselves to  a  limited set  of physically  meaningful
transformations, i.e.  between  equivalent or neighbouring filters  in
both filter sets.   We nevertheless introduced  transformations to the
($U-V$)  index in order to get colour transformations for colour
indices that  avoid  the $b$  filter, which  is the one  for which the
XMM-OM detector  could be the  most easily affected by saturation (XMM
Users' Handbook, eds.\ Dahlem   \& Shartel 1999). We  also established
various transformations in which  the $uvw1$ filter supplants the  $u$
filter.  Indeed,  contrary to the   $u$  and $U$ filters, the   $uvw1$
filter is   defined on   the  short  wavelength side of    the  Balmer
jump. This property renders  the transformations  based on $uvw1$  far
better than those based on the  neighbouring $u$ filter, i.e. they are
more linear and essentially possible on a wider temperature range. As expected (Sect.~2.1), the $uvw1$ band can be used to correct the ($U-u$) index in the critical domain where the Balmer jump is important (see Sect.~3.1.2).

The colour transformations presented below being established on
synthetic spectra, it is obvious that they should be refined through actual observations of suitable standard fields with the satellite and from the ground. The main interest in the transformations given here is that they inform us about the kind of relations that are possible, and their validity domain.

The main characteristics of the  considered colour transformations are
summarized in    Table~2.  Two of  them   are   illustrated in Figs.~2
and~3. The analytical forms  of  the transformations are given  below,
sorted  by categories.  Though some  formal error bars are smaller, we
did not indicate uncertainties smaller than 0.001 mag in the relations
given below.

\begin{figure}[htb]
\resizebox{\hsize}{!}{\rotatebox{270}{\includegraphics{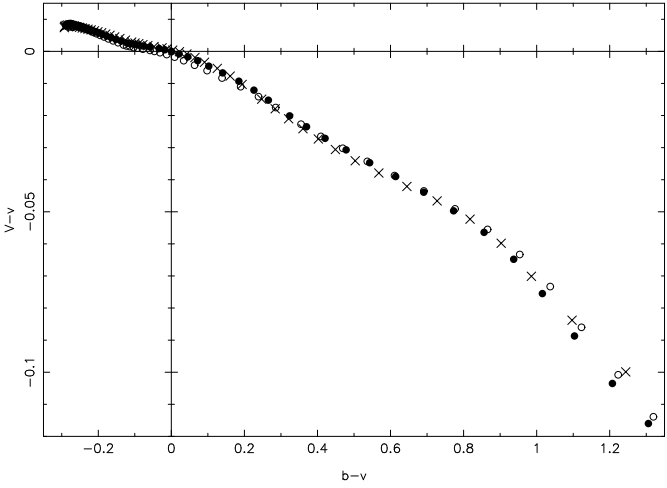}}}
\caption{The ($V-v$) vs ($b-v$) diagram. 
Filled circles are MS stars, open ones are giants and crosses 
represent metal poor stars. The lowest temperature is $4000$~K (bottom right)
and the highest is $35000$~K (upper left). The relation is monotonic and therefore the transformation is possible over the entire range.}
\end{figure}

\subsection{Solar composition main sequence stars}

\subsubsection{($U-u$) vs ($u-b$)}

\noindent\hspace{5mm}$4000$~K~$\leq~T_{\rm eff}~\leq~5500$~K (K7V - G8V)

\noindent\hspace{5mm}($u-b$)~$\in~[0.20;1.50]$\\
$U-u =  0.049 + 0.002 (u-b) - 0.097 (u-b)^2 \pm 10^{-3}$
\vspace{2 mm}

\noindent\hspace{5mm}$5500$~K~$\leq~T_{\rm eff}~\leq~10500$~K (G8V - B9V)

\noindent\hspace{5mm}($u-b$)~$\in~[-0.15;0.20]$\\
No possible transformation.
\vspace{2 mm}

\noindent\hspace{5mm}$10500$~K~$\leq~T_{\rm eff}~\leq~35000$~K (B9V - 08V)

\noindent\hspace{5mm}($u-b$)~$\in~[-1.40;-0.15]$\\
$U-u =  -0.016 - 0.268 (u-b) - 0.043 (u-b)^2 \pm 10^{-3}$

\subsubsection{($U-u$) vs ($uvw1-u$)}

\noindent\hspace{5mm}$3500$~K~$\leq~T_{\rm eff}~\leq~7250$~K (M3V - F0V)

\noindent\hspace{5mm}($uvw1-u$)~$\in~[0.35;1.75]$\\
$U-u =  -0.009 + 0.212 (uvw1-u) - 0.183 (uvw1-u)^2 \pm 4. 10^{-3}$
\vspace{2 mm}

\noindent\hspace{5mm}$7500$~K~$\leq~T_{\rm eff}~\leq~10500$~K (A8V - B9V)

\noindent\hspace{5mm}($uvw1-u$)~$\in~[-0.10;0.30]$\\
$U-u =  0.001 - 0.158 (uvw1-u) + 0.860 (uvw1-u)^2 \pm 2.10^{-3}$
\vspace{2 mm}

\noindent\hspace{5mm}$10500$~K~$\leq~T_{\rm eff}~\leq~35000$~K (B9V - O8V)

\noindent\hspace{5mm}($uvw1-u$)~$\in~[-0.75;-0.10]$\\
$U-u =  -0.019 - 0.485 (uvw1-u) - 0.128 (uvw1-u)^2 \pm 2.10^{-3}$

\subsubsection{($B-b$) vs ($b-v$)}

\noindent\hspace{5mm}$4000$~K~$\leq~T_{\rm eff}~\leq~4750$~K (K7V - K3V)

\noindent\hspace{5mm}($b-v$)~$\in~[1.00;1.30]$\\
$B-b = -0.074 - 0.010 (b-v) \pm 10^{-3}$
\vspace{2 mm}

\noindent\hspace{5mm}$4750$~K~$\leq~T_{\rm eff}~\leq~10500$~K (K3V - B9V)

\noindent\hspace{5mm}($b-v$)~$\in~[-0.05;1.00]$\\
$B-b = -0.085 (b-v) \pm 10^{-3}$
\vspace{2 mm}

\noindent\hspace{5mm}$10500$~K~$\leq~T_{\rm eff}~\leq~35000$~K (B9V - O8V)

\noindent\hspace{5mm}($b-v$)~$\in~[-0.30;-0.05]$\\
$B-b =  -0.073 (b-v) -0.065 (b-v)^2 \pm 10^{-3}$
\vspace{2 mm}

\noindent A third order fit can be applied to the whole interval:
\vspace{2 mm}

\noindent\hspace{5mm}$4750$~K~$\leq~T_{\rm eff}~\leq~35000$~K (K3V - O8V)

\noindent\hspace{5mm}($b-v$)~$\in~[-0.30;1.00]$\\
$B-b = -0.070 (b-v) - 0.040 (b-v)^2 + 0.026 (b-v)^3 \pm 10^{-3}$

\begin{figure*}[htb]
\resizebox{\hsize}{!}{\rotatebox{270}{\includegraphics{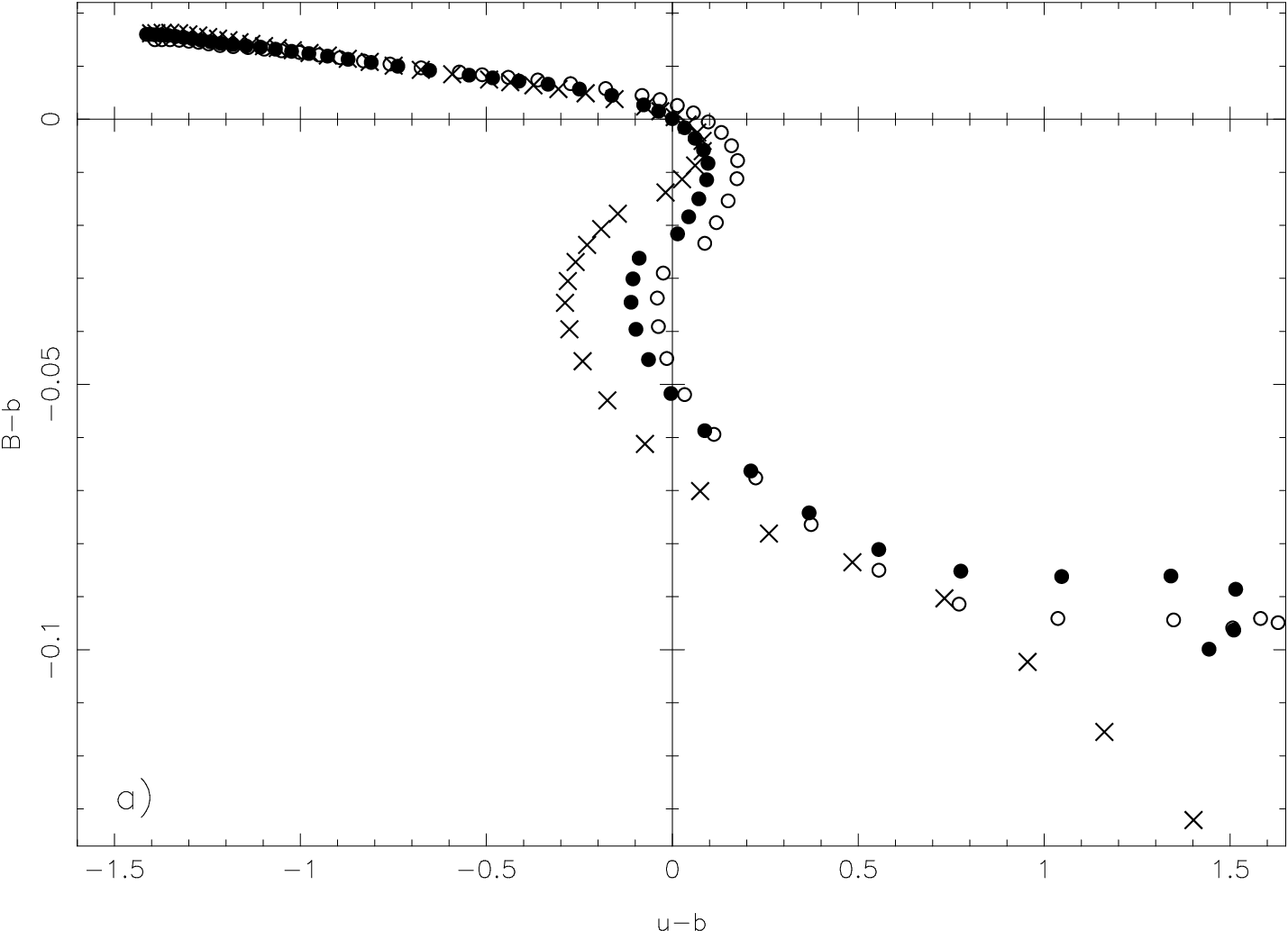}}\rotatebox{270}{\includegraphics{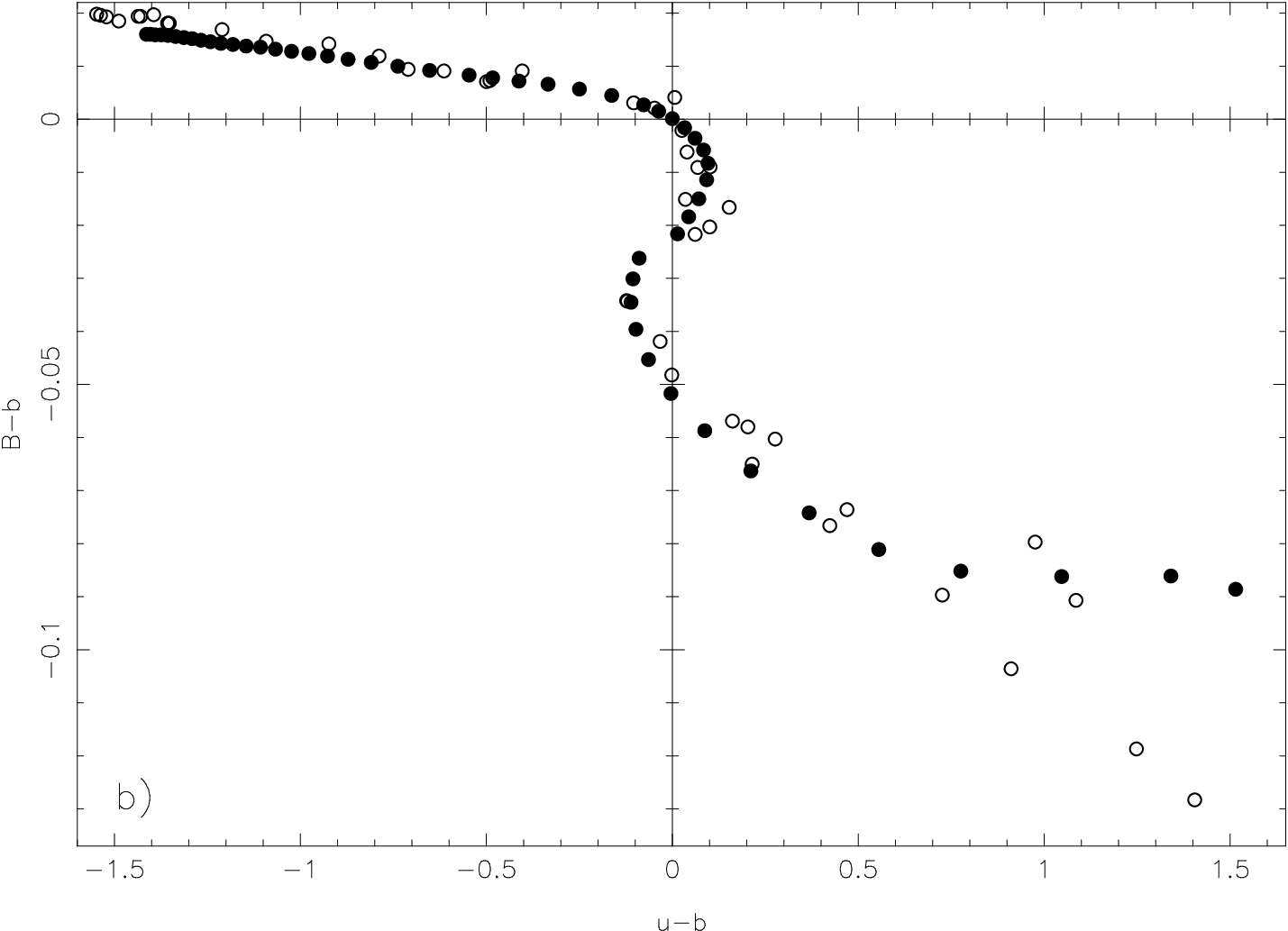}}}
\caption{{\bf{a}} ($B-b$) vs ($u-b$) diagram. 
Symbols have the same meaning as in Fig.~2. Stars have effective 
temperatures between $3500$~K and $35000$~K. This is a good illustration of a case where no transformation is possible in the middle part of the temperature domain. {\bf{b}} Same diagram for FRV MS and giant stars (open circles). Kurucz MS stars (filled circles) are given for comparison. Stars below $4000$~K have been excluded from this plot.}
\end{figure*}

\subsubsection{($B-b$) vs ($u-b$)}

\noindent\hspace{5mm}$4250$~K~$\leq~T_{\rm eff}~\leq~5500$~K (K6V - G8V)

\noindent\hspace{5mm}($u-b$)~$\in~[0.20;1.30]$\\
$B-b =  -0.056 - 0.060  (u-b) + 0.028 (u-b)^2 \pm 10^{-3}$
\vspace{2 mm}

\noindent\hspace{5mm}$5500$~K~$\leq~T_{\rm eff}~\leq~10500$~K (G8V - B9V)

\noindent\hspace{5mm}($u-b$)~$\in~[-0.15;0.20]$\\
No possible transformation.
\vspace{2 mm}

\noindent\hspace{5mm}$10500$~K~$\leq~T_{\rm eff}~\leq~35000$~K (B9V - O8V)

\noindent\hspace{5mm}($u-b$)~$\in~[-1.40;-0.15]$\\
$B-b =  0.003 - 0.009  (u-b) \pm 10^{-3}$

\subsubsection{($V-v$) vs ($b-v$)}

\noindent\hspace{5mm}$4000$~K~$\leq~T_{\rm eff}~\leq~5250$~K (K7V - K0V)

\noindent\hspace{5mm}($b-v$)~$\in~[0.85;1.30]$\\
$V-v =  0.061 - 0.136 (b-v) \pm 10^{-3}$
\vspace{2 mm}

\noindent\hspace{5mm}$5250$~K~$\leq~T_{\rm eff}~\leq~8750$~K (K0V - A3V)

\noindent\hspace{5mm}($b-v$)~$\in~[0.07;0.85]$\\
$V-v =  0.0025 - 0.068 (b-v) \pm 10^{-3}$
\vspace{2 mm}

\noindent\hspace{5mm}$8750$~K~$\leq~T_{\rm eff}~\leq~30000$~K (A3V - B0V)

\noindent\hspace{5mm}($b-v$)~$\in~[-0.30;0.07]$\\
$V-v =  -0.001 - 0.033 (b-v) \pm 10^{-3}$
\vspace{2 mm}

\noindent A second order fit can be applied to the whole interval:
\vspace{2 mm}

\noindent\hspace{5mm}$4000$~K~$\leq~T_{\rm eff}~\leq~30000$~K (K7V - B0V)

\noindent\hspace{5mm}($b-v$)~$\in~[-0.30;1.30]$\\
$V-v = -0.001 - 0.041 (b-v) - 0.035 (b-v)^2 \pm 2.10^{-3}$

\subsubsection{($U-B$) vs ($u-b$)}

\noindent\hspace{5mm}$4250$~K~$\leq~T_{\rm eff}~\leq~5500$~K (K6V - G8V)

\noindent\hspace{5mm}($u-b$)~$\in~[0.20;1.35]$\\
$U-B =  0.120 + 1.087 (u-b) - 0.135 (u-b)^2 \pm 2.10^{-3}$
\vspace{2 mm}

\noindent\hspace{5mm}$5500$~K~$\leq~T_{\rm eff}~\leq~10500$~K (G8V - B9V)

\noindent\hspace{5mm}($u-b$)~$\in~[-0.15;0.20]$\\
No possible transformation.
\vspace{2 mm}

\noindent\hspace{5mm}$10500$~K~$\leq~T-{\rm eff}~\leq~35000$~K (B9V - O8V)

\noindent\hspace{5mm}($u-b$)~$\in~[-1.4;-0.15]$\\
$U-B =  0.003 + 0.819 (u-b) \pm 5.10^{-3}$

\subsubsection{($U-B$) vs ($uvw1-b$)}

\noindent\hspace{5mm}$4000$~K~$\leq~T_{\rm eff}~\leq~6500$~K (K7V - F5V)

\noindent\hspace{5mm}($uvw1-b$)~$\in~[0.20;3.20]$\\
$U-B =  -0.292 + 0.710 (uvw1-b) - 0.051 (uvw1-b)^2 \pm 1.2\,10^{-2}$
\vspace{2 mm}

\noindent\hspace{5mm}$6500$~K~$\leq~T_{\rm eff}~\leq~8750$~K (F5V - A3V)

\noindent\hspace{5mm}($uvw1-b$)~$\in~[0.10;0.20]$\\
No possible transformation.
\vspace{2 mm}

\noindent\hspace{5mm}$8750$~K~$\leq~T_{\rm eff}~\leq~35000$~K (A3V - O8V)

\noindent\hspace{5mm}($uvw1-b$)~$\in~[-2.20;0.10]$\\
$U-B =  0.002 + 0.528 (uvw1-b) \pm 8.10^{-3}$

\subsubsection{($B-V$) vs ($b-v$)}

\noindent\hspace{5mm}$3500$~K~$\leq~T_{\rm eff}~\leq~35000$~K (M3V - O8V)

\noindent\hspace{5mm}($b-v$)~$\in~[-0.30;1.40]$\\
$B-V =  0.002 + 0.997 (b-v) \pm 9.10^{-3}$

\subsubsection{($U-V$) vs ($u-v$)}

\noindent\hspace{5mm}$4000$~K~$\leq~T_{\rm eff}~\leq~6500$~K (K7V - F5V)

\noindent\hspace{5mm}($u-v$)~$\in~[0.40;2.80]$\\
$U-V = 0.133 + 0.947 (u-v) \pm 2.3\, 10^{-2}$
\vspace{2 mm}

\noindent\hspace{5mm}$6500$~K~$\leq~T_{\rm eff}~\leq~8500$~K (F5V - A4V)

\noindent\hspace{5mm}($u-v$)~$\in~[0.20;0.40]$\\
No possible transformation, though the deviation from the 
general shape of the sequence is very slight 
(see below for further comments).
\vspace{2 mm}

\noindent\hspace{5mm}$8500$~K~$\leq~T_{\rm eff}~\leq~35000$~K (A4V - O8V)

\noindent\hspace{5mm}($u-v$)~$\in~[-1.70;0.20]$\\
$U-V =  0.008 + 0.843 (u-v) \pm 8.10^{-3}$
\vspace{2 mm}

The general shape of our synthetic main sequence is pretty 
close to a straight line in the ($U-V$) vs ($u-v$) diagram, 
so that, if precision is not critical (at most 0.1 mag), 
one can also use the following relation:
\vspace{2 mm}

\noindent\hspace{5mm}$4000$~K~$\leq~T_{\rm eff}~\leq~35000$~K (K7V - O8V)

\noindent\hspace{5mm}($u-v$)~$\in~[-1.70;2.80]$\\
$U-V =  0.100 + 0.930 (u-v) \pm 5.7\,10^{-2}$

\subsubsection{($U-V$) vs ($uvw1-v$)}

\noindent\hspace{5mm}$4000$~K~$\leq~T_{\rm eff}~\leq~7000$~K (K7V - F1V)

\noindent\hspace{5mm}($uvw1-v$)~$\in~[0.65;4.50]$\\
$U-V =  -0.201 + 0.786 (uvw1-v) - 0.029 (uvw1-v)^2 \pm 1.9\,10^{-2}$
\vspace{2 mm}

\noindent\hspace{5mm}$7000$~K~$\leq~T_{\rm eff}~\leq~7750$~K (F1V - A7V)

\noindent\hspace{5mm}($uvw1-v$)~$\in~[0.50;0.65]$\\
No possible transformation. The problem is however very slight 
and seems to be due to the Kurucz synthetic spectra rather 
than to the filters. The best here is to use the general 
relation proposed below for the whole temperature range.
\vspace{2 mm}

\noindent\hspace{5mm}$7750$~K~$\leq~T_{\rm eff}~\leq~35000$~K (A7V - O8V)

\noindent\hspace{5mm}($uvw1-v$)~$\in~[-2.45;0.50]$\\
$U-V =  0.004 + 0.577 (uvw1-v) \pm 9.10^{-3}$
\vspace{2 mm}

The same kind of considerations as for the ($U-V$) vs \\
($u-v$) case holds concerning a general relation for the whole set of 
temperatures:
\vspace{2 mm}

\noindent\hspace{5mm}$4000$~K~$\leq~T_{\rm eff}~\leq~35000$~K (K7V - O8V)

\noindent\hspace{5mm}($uvw1-v$)~$\in~[-2.45;4.50]$\\
$U-V =  0.023 + 0.596 (uvw1-v) \pm 4.10^{-2}$

\subsection{Giants}

Whenever  a colour  transformation is  possible   for MS  stars, it is
generally  also satisfactorily obeyed by  giant stars.  The difference
essentially lies in   the validity     range  of    the   colour
transformations. Each time a forbidden zone appears in the temperature
domain (i.e. each time there is a zero in Table~2), the transformation
relative to  the hottest stars remains valid  for giant  stars down to
$\sim 250$~K  cooler than the lower temperature   bound defined for MS
stars.   Symmetrically,  the transformation  relative to  the  coolest
giant stars  is  only valid from  stars $\sim  250$~K cooler  than the
upper bound of the  temperature interval defined for the corresponding
MS star colour transformation (see Fig.~3a).

When colour transformations are possible for MS stars 
on the whole temperature domain, they are generally valid for 
giant stars as well (Fig.~2).

\subsection{Metal poor stars}

In this case, the situation is worse. Stars still obey the same 
relations as MS stars when colour transformations exist on the 
whole temperature domain. Even in other cases, the hottest stars 
still obey the same relations, but the range on which these 
transformations remain valid is now considerably diminished: 
the ``hot'' relations now hold only for $T_{\rm eff}~\geq~11500$~K 
and the ``cool'' ones for $4750$~K~$\geq~T_{\rm eff}~\geq~4000$~K. 
Cool stars sometimes require different relations. 
This is true for the ($B-b$) vs ($u-b$) as well as for 
both ($U-u$) colour transformations (Fig.~3a).

\subsection{Comparison with observed spectra}
As a validity check,  all colour transformations presented  above were
also  derived on the basis of  the  mainly-observed FRV spectra.  When
excluding   stars cooler   than   $4000$~K, the    comparison  between
transformations established through  Kurucz models and through  FRV MS
and  giant star spectra is excellent,  as illustrated  in Fig.~3b. The
latter ones are of course  slightly more dispersed, but the dispersion
is fully comparable to   what could be  expected  on the basis of  the
transformations established in Sect.~3.2. Very small differences occur
in some particular colour indices and  are quite marginal: the ($B-b$)
colour  index discriminates between the FRV  MS and  giant stars, but
this  only happens below $5000$~K; ($V-v$)  is decreased by
$\sim~0.01$~ mag when calculated on the FRV spectra rather than on the
Kurucz spectra; the Kurucz and FRV ($U-B$)  and ($U-V$) colour indices
are slightly  discrepant below $5000$~K,  as  are the  ($U-u$) and the
($uvw1-u$) colour indices below $7000$~K.

As a  conclusion,  before the full   inflight calibration on  standard
fields is performed   and reduced, we recommend  to  use the relations
given above to analyse the first data provided by the XMM-OM.

\begin{figure}[htb]
\resizebox{\hsize}{!}{\rotatebox{270}{\includegraphics{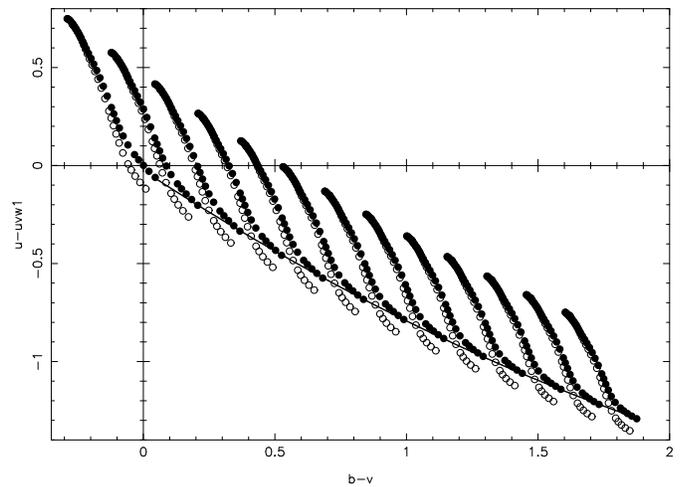}}}
\caption{The ($u-uvw1$) vs ($b-v$) colour diagram. 
Filled circles are MS stars, open ones are giants. 
Each ``temperature track'' (almost vertical)
runs from $9000$~K (bottom) to $35000$~K (top). 
The reddening runs from $A_{\rm V}~=~0.$ (leftmost track) to $A_{\rm V}~=~6.$
(rightmost track) by steps of 0.5.}
\end{figure}

\section{Reddening and temperature determination}

In order to discriminate between stars  having different reddening and
temperature characteristics, a  combination of colour  indices must be
found such that temperature tracks (i.e. curves of constant reddening)
do not cross each other in the  related colour diagram.  In the XMM-OM
photometric  system, we found only  one  fully suitable  pair of  such
colour  indices:  ($u-uvw1$)  and ($b-v$).   The  corresponding colour
diagram is  shown  in Fig.~4. This figure    is representative of  the
weakest reddenings, but the diagram keeps its properties up to $A_{\rm
V}~\sim~13$. Only a slight adaptation of  the lowest temperature bound
($9000$~K~$\rightarrow~11000$~K) is necessary.  We  did not plot stars
below $9000$~K  in Fig.~4 because the  temperature tracks  of MS stars
begin to overlay at this temperature, so that  nothing can be said for
lower temperatures. Uncertainties on the photometry will anyhow probably
hamper  any  temperature  determination   below  $10000$~K  (a 0.05  mag
photometric  error can bring any  $10000$~K  star on the $9000$~K star
locus).   Above that value,     the   precision on   the   temperature
determination depends on the temperature itself, but is typically of a
few thousand Kelvins for  colour indices accurate  to 0.05 mag and for
stars in the middle of the temperature  domain. Such accuracy allows a
reddening   determination   with   a  precision of    about $\sim~0.3$
magnitudes in  $A_{\rm V}$  at  any but  the ``coolest''  temperatures
($<~11000$~K).

\begin{figure*}[thb]
\resizebox{\hsize}{!}{\rotatebox{270}{\includegraphics{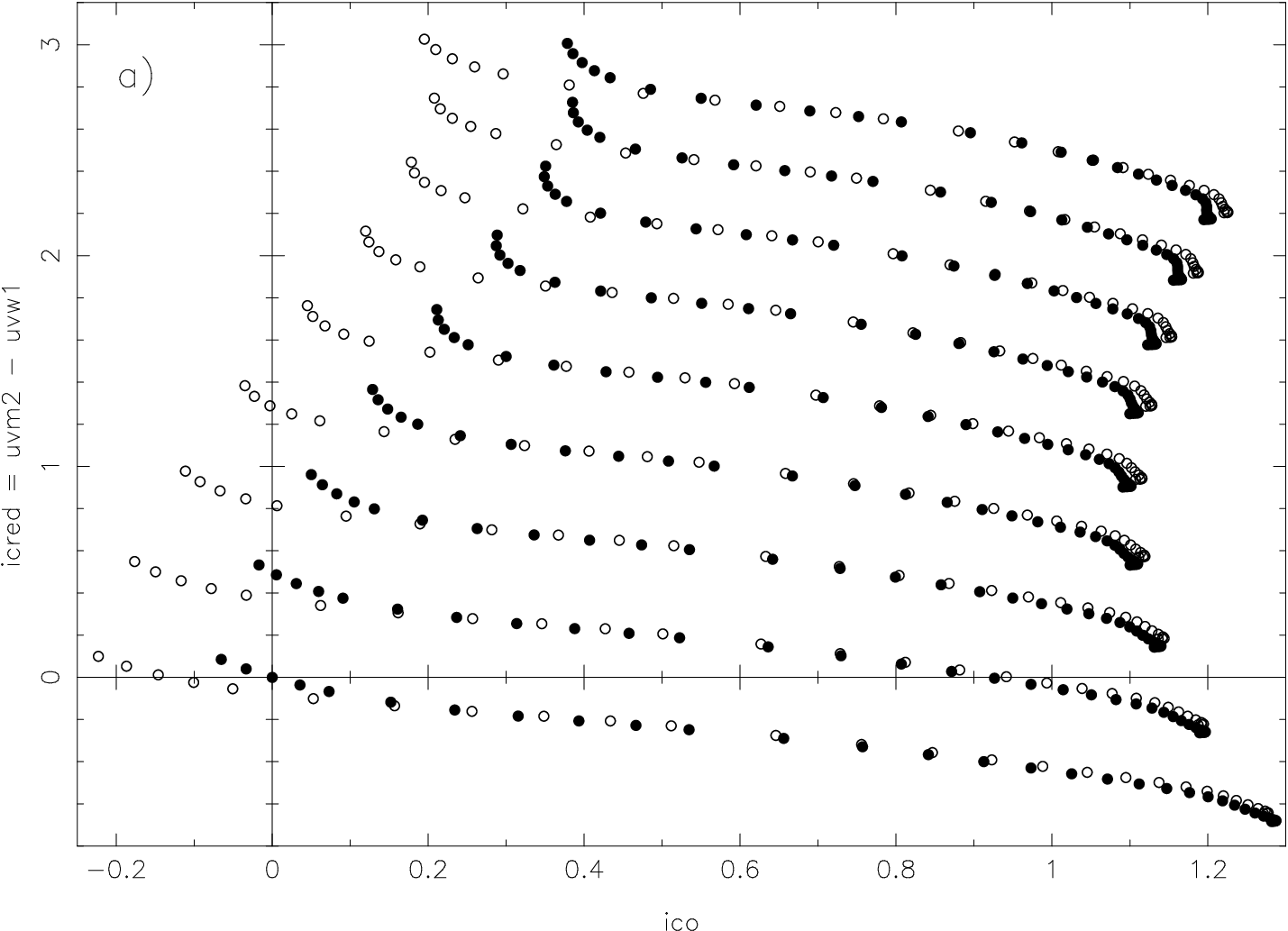}}\rotatebox{270}{\includegraphics{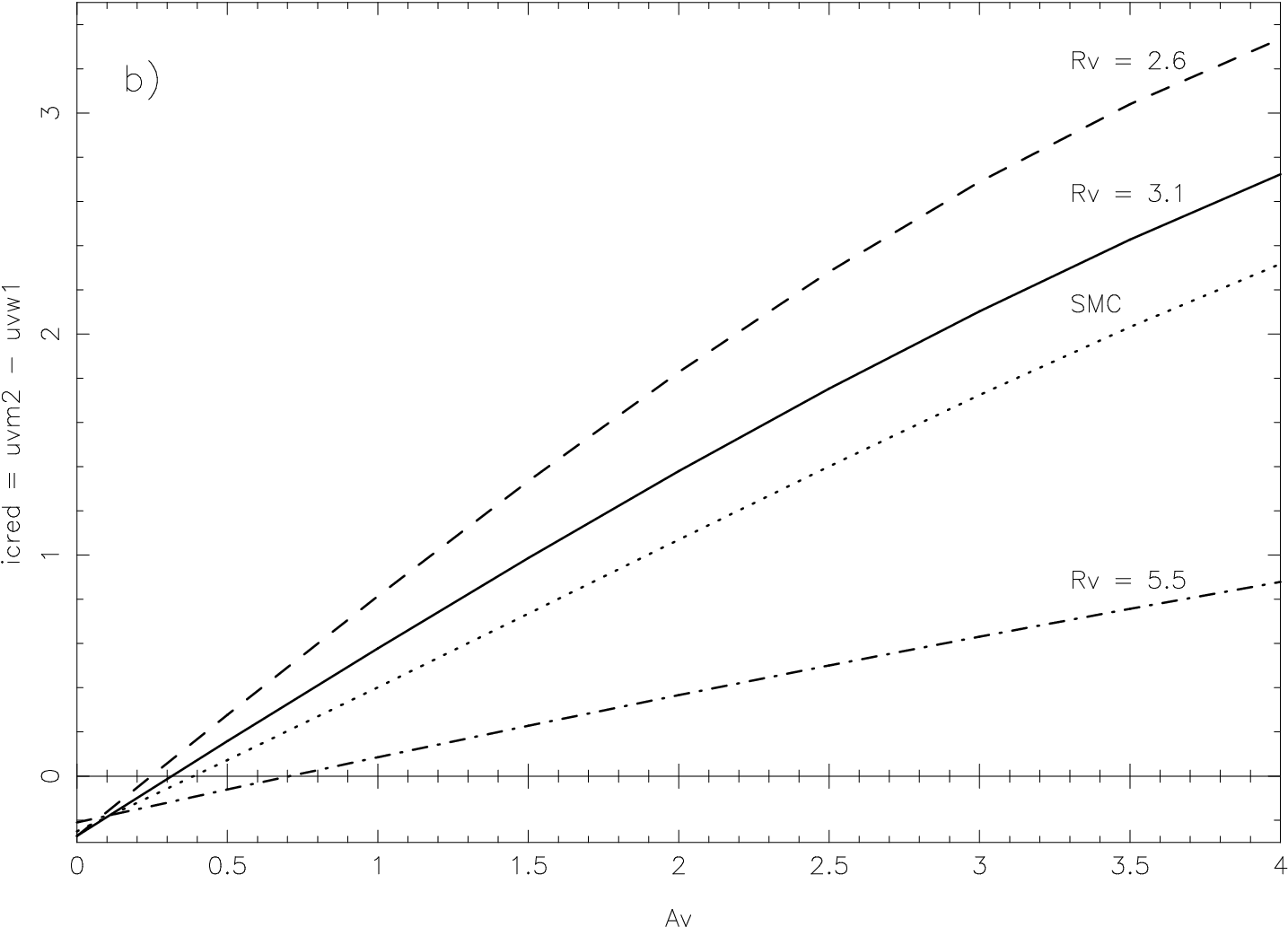}}}
\caption{{\bf{a}} The $ic_{\rm red}$ vs $ic_{\rm o}$ colour diagram. Filled circles are MS stars, open ones are giants. Each temperature track runs from 9000~K (left) to 35000~K (right). The reddening runs from $A_{\rm V}~=~0.$ (bottom track) to $A_{\rm V}~=~4.0$ (upper track). 
{\bf{b}} Dependency of the $A_{\rm V}~=~A_{\rm V}(ic_{\rm o},ic_{\rm red})$ function on the reddening law. This diagram shows a section at $ic_{\rm 0}~=~0.6$ in Fig.~5a ($R_{\rm V}~=~3.1$, full line) and in each of the equivalent diagrams established for $R_{\rm V}~=~2.6$ (dashed line), $R_{\rm V}~=~5.5$ (dash-dotted line) and for the SMC reddening law (dotted line).
}
\end{figure*}

The situation is slightly worse when one includes giant stars as well
since the temperature tracks for the giants at a given
reddening cross those of the less  reddened MS stars. Even worse, they
pass below the locus of  the $9000$~K MS stars at  $\sim~10500$~K.
As any place below this line can be occupied by stars with various
combinations of temperature and reddening, the domain in which a
temperature determination is possible for giants is restricted to
stars above  $\sim11000$~K.  As one does not know {\it a priori}
whether or not a star is a giant, this of course also sheds some
uncertainty on the temperature and reddening determination for MS stars
below $\sim12-13000$~K,  where temperature tracks for MS stars and
giants begin to significantly differ from each other.

Other discriminant colour  indices than  those of   Fig.~4 exist  in the
system,  but  their combination   with  other ones only allows to
determine the temperature and the reddening  of the observed stars on much  
more  restricted ranges of reddening  and/or temperature.  
One   can nevertheless  design   another,  independent,
reddening determination technique. In Fig.~4, reddening and temperature
influence both colour indices plotted on the axes.  As  we will see now,  one
can  nearly decouple   these parameters and  obtain  better precision on the
reddening.  To do so,  we need   to define  a  colour  index that  is
independent of reddening and another that  is proportional to it.  The
former, that we will call $ic_{\rm 0}$, is defined as
$$ic_{\rm 0}   =  (uvw2-uvw1) - \frac{E(uvw2-uvw1)}{E({u-uvw1)}}  (u-uvw1)$$
where $E(x-y)$ stands for the colour excess of the ($x-y$)
colour index. $ic_{\rm o}$ will remain a reddening free colour index  
as long as the colour excess ratio remains constant. 
This property is verified for
stars with  $T_{\rm eff}~\geq~9\,000$~K and $A_{\rm V}~\leq~5.5$. 
Nevertheless, since $A_{\rm 2000}~\sim~3~A_{\rm V}$, four magnitudes of absorption in $V$ correspond to more than ten magnitudes of absorption in the $uvm2$ and $uvw2$ filters. Hence, from now on, we will only consider the $A_{\rm V}~\leq~4$ domain. The mean $E(uvw2-uvw1)/E(u-uvw1)$ colour excess ratio over this range is $-2.83$ ($\pm\,0.27$). It is worth to note that  $(uvw2-uvw1)$ and $(u-uvw1)$ are, with $(u-b)$ and $(b-v)$, the only pairs of colour indices allowing the definition of a reddening free index over a reasonable range of stellar parameters.

In order to define the reddening dependent index, that we will call
$ic_{\rm red}$, the most obvious choice is the empirical
$$ic_{\rm red}~=~uvm2~-~(uvw1~+~uvw2)\,/\,2.$$
This index could be used but it is not really independent from $ic_{\rm 0}$ and its reddening dependence can be ameliorated. Indeed, though the $uvm2$ filter stands on the $2175$~\AA\ absorption bump, the $uvw1$ and $uvw2$ filters are not symmetric with respect to it, and of more importance, the $uvw2$ filter is significantly affected by the $2175$ \AA\ absorption bump too. To refine the choice, we explored a large number of UV magnitude combinations and were finally brought to the conclusion that the best choice in terms of simplicity and dynamics of the index (and hence in terms of accuracy) is 
$$ic_{\rm red}~=~uvm2~-~uvw1$$ 

\begin{figure*}[tbh]
\resizebox{\hsize}{!}{\rotatebox{270}{\includegraphics{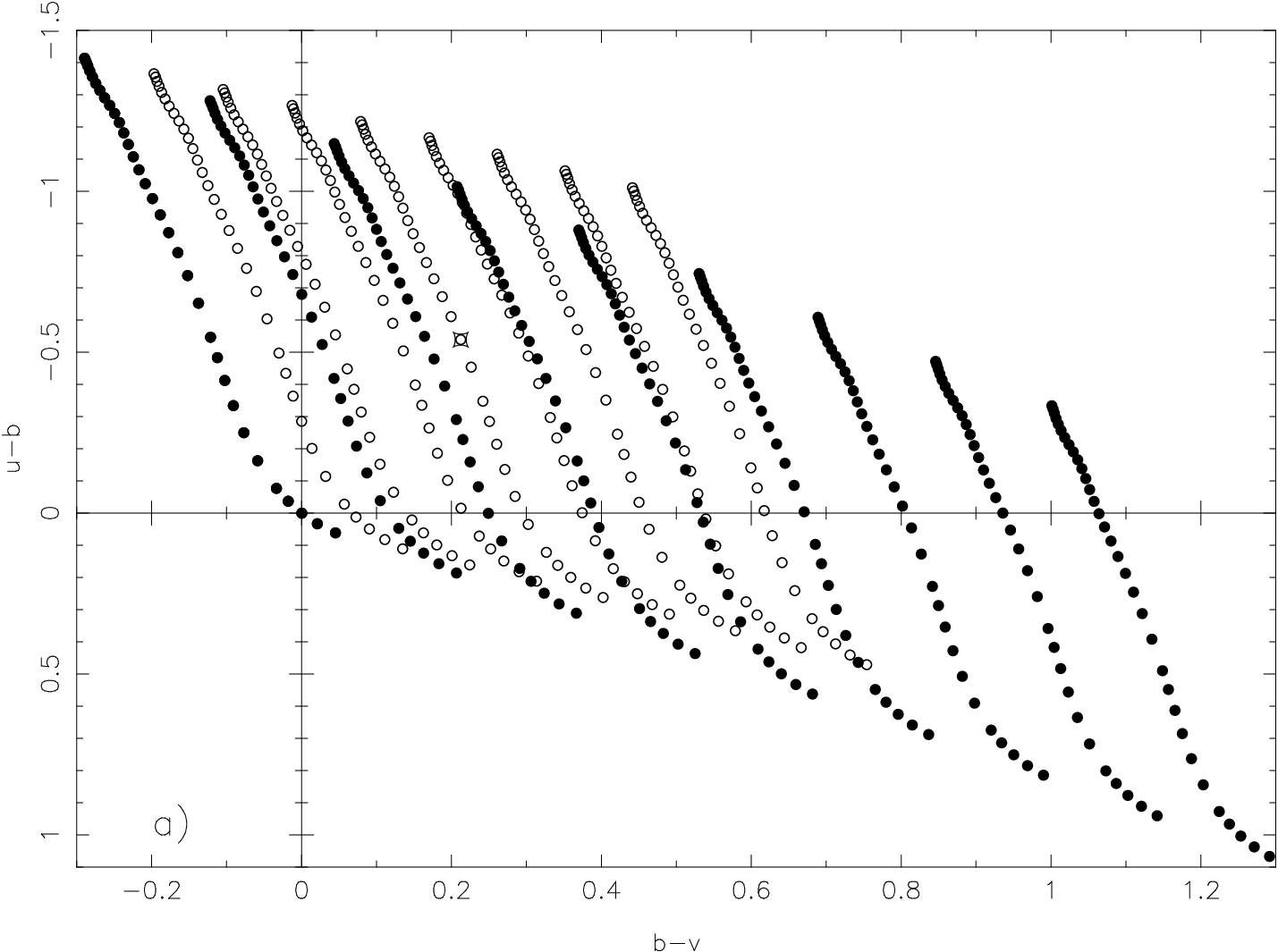}}\rotatebox{270}{\includegraphics{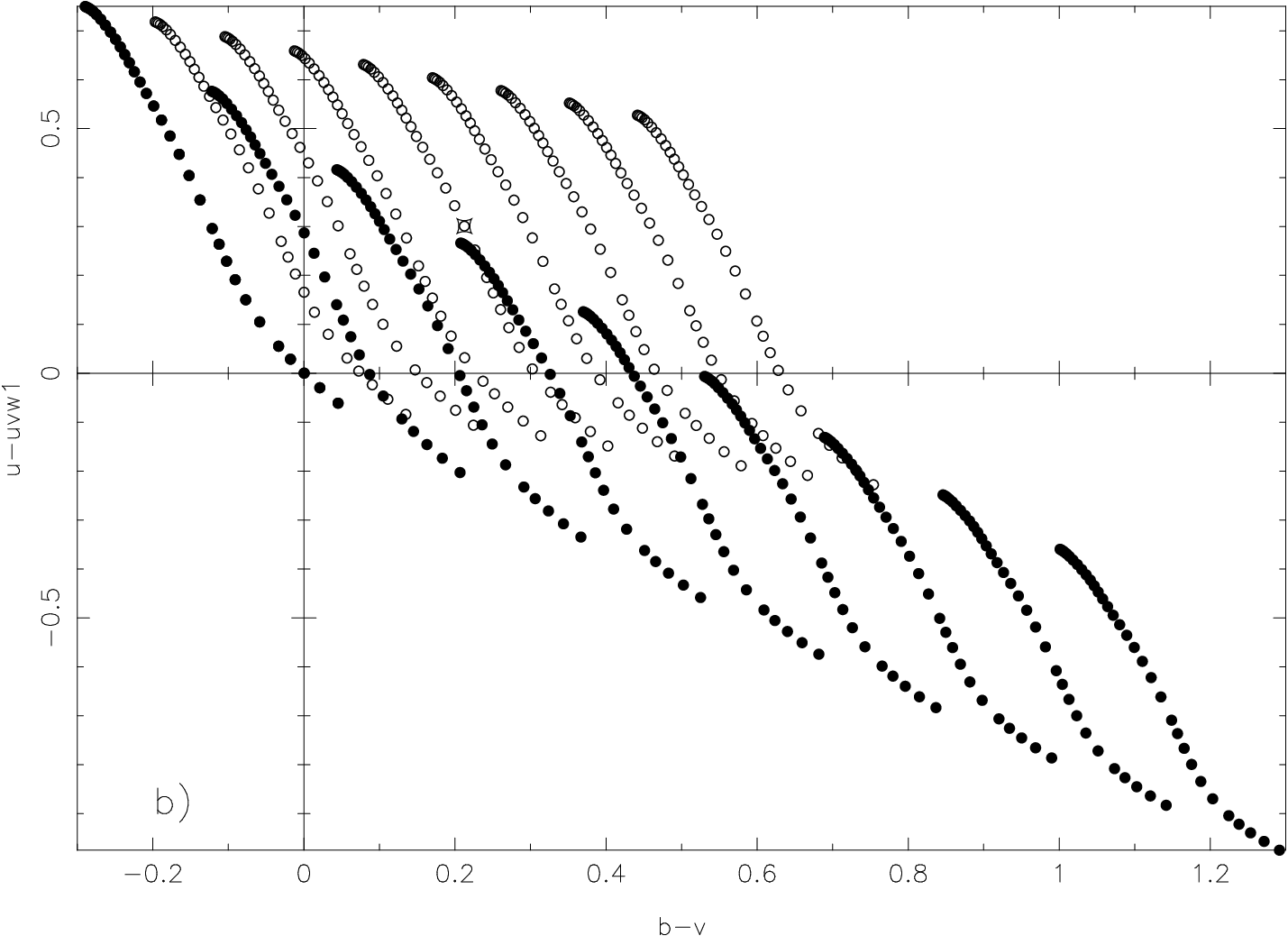}}}
\caption{{\bf{a}} The ($u-b$) vs ($b-v$) colour diagram 
for MS stars reddened according to Cardelli et al. (1989) reddening law with $R_{\rm V}~=~3.1$ (filled circles) and  $R_{\rm V}~=~5.5$ (open circles). Each temperature track runs from 9000~K (bottom) to 35000~K (top). The reddening runs from $A_{\rm V}~=~0.$ (leftmost track) to $A_{\rm V}~=~4.0$ (rightmost track). The additional diamond symbol points to the star with $T_{\rm eff}~=~15000~K$, $A_{\rm V}~=~2.$ and $R_{\rm V}~=~5.5$ (see text). 
{\bf{b}} The ($u-uvw1$) vs ($b-v$) colour diagram with the same conventions. 
}
\end{figure*}

There is   in  fact a wide variety   of
reddening-dependent indices  in the  XMM-OM system  and, to give just
another  very simple   one,  ($uvw2-uvw1$)  is   nearly as   good   as
($uvm2-uvw1$). Even ($b-v$)   could  be used as  reddening-dependent
index,  so that the whole treatment  that we carry  on  here on the UV
filters could  be performed on the  $ubv$ filters as well. Nevertheless, it is important to note that, although the reddening-free  index based on  the optical filters is  much better than the one
based on the ultraviolet filters ($E(u-b)/E(b-v)~=~0.80~\pm~0.02$; compare the error bar with the UV case), the dynamic of the ($b-v$) index
is smaller  than the  one  of  the  ($uvm2-uvw1$) index. Consequently, at
equivalent accuracy on the photometry, the reddening determined thanks
to the UV filters  will be more accurate  than the one obtained on the
basis  of the optical photometry (except for    temperatures lower than
10000~K  or  higher than 25000~K). If  UV   and optical photometries are
available, reddening determinations  in both domains  should of course
be used as a quality check. Indeed, as we will see in Sect.~5, the simultaneous use of visible and UV data allows a control of the consistency of the adopted reddening law.

Although lots of reddening dependent indices  exist, none is perfectly
independent  from temperature and from  $ic_{\rm 0}$ so that no simple
$A_{\rm V}~=~A_{\rm V}(ic_{\rm red})$ relation can be drawn.  Instead,
one  has  to   perform  a  bi-dimensional   fit  over   the  ($ic_{\rm
0}$,$ic_{\rm red}$) plane  (Fig.~5a) to  get the $A_{\rm   V}~=~A_{\rm
V}(ic_{\rm 0},ic_{\rm   red})$ relation.  We do   not present this fit
here since its detailed analytical form can  only be usefully obtained
through actual  in-orbit   satellite  calibrations.   The    reddening
dependency  of $ic_{\rm red}$ is  illustrated  in Figs.~5a and~5b.  The
solid line in Fig.~5b represents a section of the colour diagram shown
in Fig.~5a at $ic_{\rm 0}~=~0.6$.

This reddening  determination method, based on  a reddening free and a
reddening dependent index, is more  accurate than what can be expected
with the procedure outlined in Fig.~4 since error bars of 0.2~and 0.05~mag on \ico\ and \icr\ respectively lead to  $\leq~0.15$ mag uncertainty  on $A_{\rm V}$.  On
the other hand, one can  see by comparing  Fig.~5a and Fig.~4 that the
uncertainties  on  $A_{\rm V}$, due   to  the fact that the  temperature
tracks for MS and giant stars diverge at  the coolest temperatures in
the $(u-uvw1)$~vs~$(b-v)$ colour  diagram, are partly removed here  for
the weakest reddenings.    Nevertheless, the colour  diagram  shown in
Fig.~4  remains  necessary since  the  $A_{\rm V}~=~A_{\rm  V}(ic_{\rm
0},ic_{\rm red})$ fit  is not valid for all  temperatures, so that  we
need an independent determination of temperature, precisely allowed by
Fig.~4.

\section{Non-standard reddening laws}

The reddening determination method presented in the previous section does not take into account the possibility of a non-standard reddening law. We will now address the following question: can we determine whether or not the reddening of a star observed with the XMM-OM is anomalous ? To answer this, we integrated the same spectra with alternative values of $R_{\rm V}$. Instead of $R_{\rm V}~=~3.1$, ---the standard value for diffuse interstellar medium and average value for the LMC---, we used $R_{\rm V}~=~2.6$ and $R_{\rm V}~=~5.5$ 
(observed by Cardelli et al. 1989, in the direction of HD\,204827 and HD\,37022 respectively). We also considered the peculiar UV absorption law established by Pr\'evot et al. (1984) for the SMC. This reddening law does not show any absorption bump around $2175$~\AA. It is specific to the ultraviolet and was thus only applied to the UV filters.

Figs.~6a and~6b show the resulting $(u-b)$~vs~$(b-v)$ 
and $(u-uvw1)$~vs~$(b-v)$ colour diagrams in the $R_{\rm V}~=~5.5$ case. Let us consider a $15000$~K star affected by 2 magnitudes of absorption (in $V$) with a reddening law characterized by such a high value of $R_{\rm V}$ and let us see what will happen if we try to interpret the observed colours assuming a classical reddening law with $R_{\rm V}~=~3.1$. In the $(u-b)$~vs~$(b-v)$ diagram, the colours mimic those of a slightly hotter star, affected by hardly more than one magnitude of absorption but in $(u-uvw1)$~vs~$(b-v)$, 
the star falls out of the ``authorized'' range of colours. This discrepancy will clearly reveal that the assumption was wrong and that the star suffers from a peculiar absorption law. Fig.~6 thus reveals that $(u-uvw1)$, $(b-v)$ and $(u-b)$ can be used to determine $T_{\rm eff}$, $A_{\rm V}$ and $R_{\rm V}$ just as Fig.~4 shows that $(u-uvw1)$~vs~$(b-v)$ can be used to determine $T_{\rm eff}$ and $A_{\rm V}$ once $R_{\rm V}$ is known.

Of course, peculiar absorption laws also influence \ico\ and \icr. As expected from Fig.~6, the colour excess ratios are reddening law dependent. Considering $0~\leq~A_{\rm V}~\leq~4.$, we found the mean value of ${E(uvw2-uvw1)}/{E(u-uvw1)}$ to vary between $-2.83$ and $-3.68$ for $R_{\rm V}~\in~[2.6;5.5]$ (the value for the Pr\'evot et al. (1984) reddening law is $-3.04$). Consequently, the appropriate (\ico,\icr) has to be used once $R_{\rm V}$ has been determined. Fig.~5b shows the influence of the reddening law variations on the (\ico,\icr) diagram through comparison between 
$ic_{\rm 0}~=~0.6$ sections performed in the various 
realizations of that diagram.

As a conclusion, regarding the reddening determination, 
the main advantages of the XMM-OM UV filters are that
\begin{itemize}
\item at equivalent photometric accuracy, they allow a better determination of the amount of interstellar extinction than what is achievable with the optical filters only;
\item they allow discrimination between standard and non-standard extinction laws; this is of primary importance since interpretation of X-ray data requires knowledge of the column density along the line of sight, what is often estimated through reddening measurements (e.g. Bohlin et al. 1978).
\end{itemize}

\section{The XMM-OM multicolour space and quasars}

In the present section, we investigate the ability of the
XMM-OM photometric system to segregate quasars from stars
on the basis of their colours in the multicolour space
definable from the set of the different filters used.
Therefore, we integrated both the stellar spectra discussed
in Sect.~2.3 and the average quasar spectra discussed in
Sect.~2.4. The latter ones have been considered at redshifts from
0.0 to 4.4 by steps of 0.1. We start our analysis with the
XMM-OM version of the classical ($U-B$) vs ($B-V$) colour diagram.

\subsection{The basic ($u-b$) vs ($b-v$) diagram}

The  ($U-B$) vs  ($B-V$) colour diagram  is probably  one of  the most
widely  used  in  astronomy.   Therefore, Fig.~7 presents   the XMM-OM
version (($u-b$) vs  ($b-v$)) of this  diagram.   Fortunately, no huge
difference appears between the two versions and the XMM-OM colour diagram
retains most of the properties of its classical counterpart. 

\begin{figure}[htb]
\centerline{\resizebox{7.3cm}{!}{\rotatebox{270}{\includegraphics{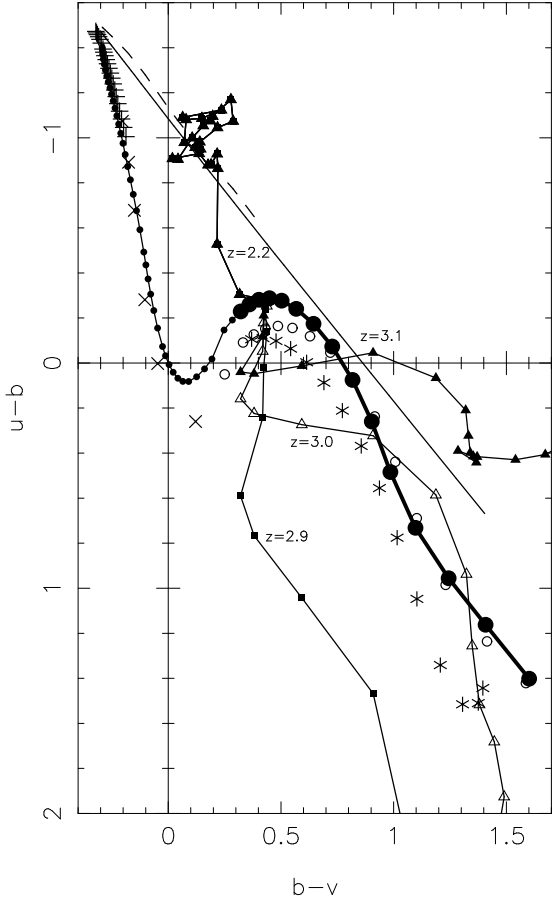}}}}
\caption{The ($u-b$) vs    ($b-v$)  colour diagram.  The    halo  main
sequence  stars  are represented by  filled circles;  the part  of the
sequence populated in  our Galaxy   is in  bold character.   The  halo
giants are represented  by open circles  whereas  the cool  end of the
disk main  sequence is described by  asterisks. The crosses  stand for
the HB BA stars whereas the plus signs deal with the  sd OB stars. The
continuous line  represents the black body  locus and  the dashed line
the positions   of   the degenerate stars    as  derived from Koester's
models.  Model~A average quasar    spectrum  has its redshift    track
punctuated with filled triangles every 0.1 in redshift whereas model~B
is  marked  out by open  triangles  and model~C  by filled squares. The 
highest plotted redshifts are 4.1, 3.7 and 3.1 for the quasars of models 
A, B and C respectively.}
\end{figure}

The locus of the theoretical halo main sequence  stars is given (filled circles). The classical potential-well 
shape of the curve outlining the effect
of the Balmer continuum is clearly visible. The part below the turn-off represents the stars that are still on the
main sequence in the halo of our Galaxy; it  is represented by a bold
line. This bold line is the  locus of the  majority of the field stars
which are the  objects  against  which  we have to  perform  the basic
discrimination when looking for quasars.  The locus  of cool giants is
also given   (open  circles) and is very   similar  to the  previously
discussed one.   On the other hand,  the  disk main sequence (spectral
types FGKM), which could also provide  some confusion, is visible below
the   halo  main sequence  (asterisks).    The HB stars (crosses) have
colours very similar to the hot part (astronomically non populated) of
the halo  main sequence.  This is also  true  for the  sd\,OB stars (plus
signs).  These  objects are classical   contaminants of the samples  of
quasar candidates selected on  the basis of  the $U/B$ excess. Another
contaminant are  the degenerate stars.   The black  body line  is also
given in Fig.~7 and the Koester's models whose spectrum we integrated 
fall close to this line. 
Fig.~7  also  exhibits the track of our
average quasars as a function of redshift.  For redshifts $z$ $\leq$ 2.1,
the average   quasars    are    wandering    around $u-b$~=~$-1.0$    and
$b-v$~=~0.15. This is 0.2 magnitude bluer in $u-b$ than in the case of
the standard system  (see e.g.\ Fig.~6  of  Cristiani \& Vio  1990 and
Fig.~9 of Moreau \&  Reboul 1995). This ability  to better  detect the
bluer   objects   is essentially   due   to  the   comparatively  high
transmission of the $u$ filter below 3200 \AA\ (see Fig.~1b and Sect.~2.1).

The wandering around the mean place  is essentially due to the various
emission lines entering and  going  out of  the different  filters but
also partly to  the particular shapes  of the top of the  transmission
curves. The region  around this mean place  is also where one can find
degenerate stars (with typical $T_{\rm eff} \sim 15000$~K)
and the  presently investigated colour combination is
not  useful   in  discriminating between   both   kind of objects.  At
$z$~=~2.2, the quasars have left their low-redshift location to  move
almost parallel to the ($u-b$) axis. This is due  to the fact that the
Ly$\alpha$ emission  line is leaving  the $u$ filter to enter the $b$
one  and  is progressively  replaced by the  Ly$\alpha$  forest.  This
occurs 0.07 earlier in redshift than in the Johnson standard system.

The tracks of the average quasars are also
given for higher redshifts. 
It should however be clearly stated that for redshifts
$z~\geq~2.9$, both the $u$ and the $b$ filters are mainly sampling
the Ly$\alpha$ forest and the related quasar location in the bidimensional 
(2D) colour diagram is highly model dependent. It is interesting
to notice that the model~B spectra tend to follow the
stellar locus whereas model~C quasars stay bluer in
$b-v$. In any case, the discrimination is essentially
possible for low-redshift ($z~\leq~2.2$) quasars and for
high-redshift weakly or strongly absorbed objects.

Fig.~8 gives the distance (in  this two-dimensional space) between the
model~A quasar and  the stellar locus  as a  function of redshift  (by
steps of 0.1, filled circles).  The  stellar locus adopted here is the
whole halo main sequence. Except  for degenerate stars, this locus can
be  considered   as representative  of  most of   the  other potential
contaminants.

Fig.~8 clearly demonstrates the ability to discriminate
between non-degenerate stars and quasars with redshifts
$z~\leq~2.2$. In the range $z~\in~[2.2, \sim 3.]$ (or more), quasars have 
$ubv$ colours very similar to stars and the discrimination power will
only improve through the use of additional filters. 
The apparent improvement for redshifts in the range $z~\in~[3.3,3.6]$ is not present for model~B quasars: this pinpoints 
the model dependent character of this particular result.

At $z~>~3.8$, all three filters are essentially in the 
Ly$\alpha$ forest and the increase in distance
is indicative of a potential discrimination but the latter
is bound to be highly dependent on the particular
realization of the distribution of the Ly$\alpha$
absorbers both in redshift and in density. The completeness
of the related sample will be hard to ascertain.

\begin{figure}[htb]
\resizebox{\hsize}{!}{\rotatebox{270}{\includegraphics{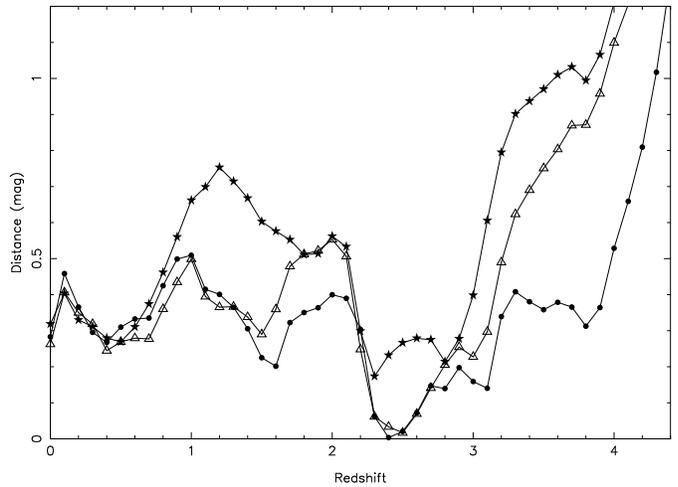}}}
\caption{The reduced distance  between the average  model~A quasar and
the stellar  locus  as a   function   of redshift, by steps of 0.1.   
The  first  curve
(circles) represents the two-dimensional distance in the 2D ($u-b$) vs
($b-v$)  colour  diagram  as  given  in  Fig.~7.   The  second   curve
(triangles) represents        $\sqrt{2}  /   \sqrt{3}$      times  the
three-dimensional distance in the 3D ($uvw1-u$)  vs ($u-b$) vs ($b-v$)
colour diagram. The   third  curve  (stars)  gives $\sqrt{2}  /
\sqrt{4}$ times the  four-dimensional distance in the 4D ($uvw2-uvw1$)
vs ($uvw1-u$) vs ($u-b$) vs ($b-v$) colour diagram.}
\end{figure}

\subsection{Adding $uvw1$}

Our simulations concerning the $uvw1$ filter clearly
indicate that this filter is roughly as sensitive to the Balmer continuum 
(not to confuse with the Balmer jump) as the $u$ filter. 
The ($uvw1-b$) colour index of stars has a behaviour very similar
to the ($u-b$) one and the locus of stars in a 
($uvw1-b$) vs ($b-v$) diagram is very reminiscent of 
Fig.~7. The ($uvw1-u$) colour index is much less sensitive
to the Balmer continuum. It is interesting to notice that
in a 2D ($uvw1-u$) vs ($uvw1-v$) colour diagram, the quasars
with $z~\leq~1.5$ are perfectly superimposed on the locus of stars.
Therefore, this combination is not interesting for
low-redshift quasars but quasars with redshifts between 1.6 and 2.1
are moving away from the stellar locus in the same diagram.
This is essentially due to the Ly$\alpha$ emission line 
leaving $uvw1$ for the $u$ filter and to the Ly$\alpha$
forest becoming dominant in $uvw1$. Nevertheless, full exploitation of this phenomenon requires observations in the $b$ filter. This is particularly striking in the ($uvw1-u$)~vs~($u-b$) colour diagram (not shown here).

\begin{figure}[htb]
\resizebox{\hsize}{!}{\rotatebox{270}{\includegraphics{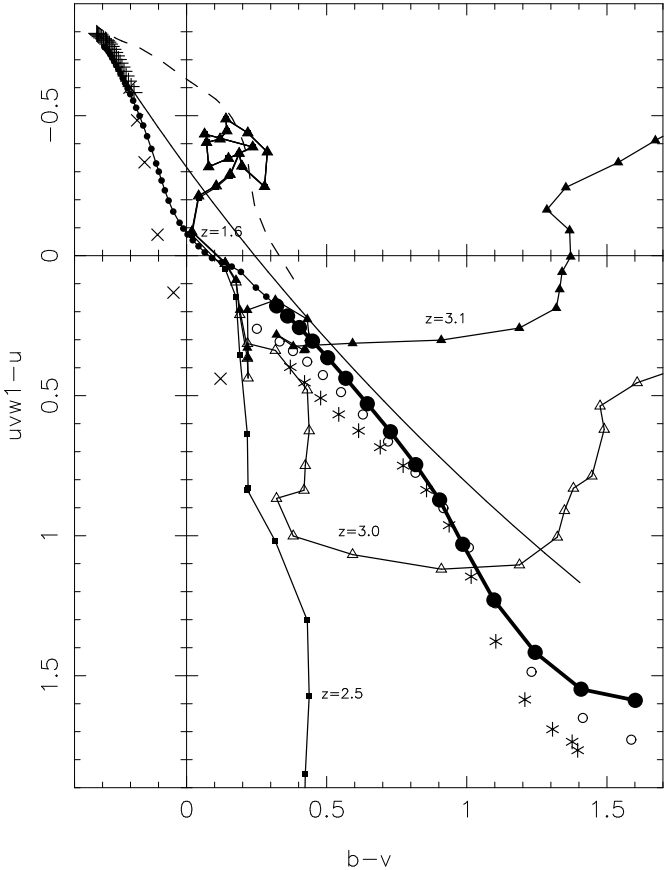}}}
\caption{The ($uvw1-u$) vs ($b-v$) colour diagram. The
symbols have the same meaning as those used for Fig.~7. 
The highest plotted redshifts are 4.1, 3.9 and 2.6 for the quasars of 
models A, B and C respectively.}
\end{figure}

Fig.~9 gives the 2D ($uvw1-u$) vs ($b-v$) colour diagram.
At redshifts $z~\leq~1.5$, quasars are wandering around
$uvw1-u$~=~$-0.4$ and $b-v$ = 0.15. This is slightly aside
the non-degenerate stellar locus and the $uvw1$ filter
contributes, although weakly, to the star-quasar separation.
For redshifts $1.6~\leq~z~\leq~3.0$, the average quasar joins the
stellar locus in this 2D diagram of Fig.~9 but it is known
to deviate from the stellar locus in the ($uvw1-u$) vs ($uvw1-v$) colour diagram for $1.6~\leq~z~\leq~2.1$. In Fig.~8 is also given the
distance between the quasar and the stellar locus 
in the three-dimensional space
($uvw1-u$) vs ($u-b$) vs ($b-v$). Increasing the number of dimensions
of the space always brings an increase of the distance between
objects, although the effect is purely geometrical. To test
whether or not the added filter brings a strategical contribution
due to its location in the wavelength domain, one
has to compare the distances reduced to the lower dimension space.
Therefore, in Fig.~8, we compare the two-dimensional true distance
(in ($u-b$) vs ($b-v$)) to the reduced~3D distance which is the
three-dimensional true distance (in ($uvw1-u$) vs ($u-b$) vs ($b-v$))
multiplied by a $\sqrt{2} / \sqrt{3}$ factor.

From Fig.8,  it is absolutely clear  that the main contribution of the
use of $uvw1$ to the discriminating  power of the XMM-OM photometry is
essentially located  at  the redshift range 1.6  to  2.1.   It is also
interesting to  notice  that  low-redshift  quasars are wandering   in
Fig.~9 slightly  aside the black  body  line.  However, the degenerate
stars do not follow the black body  locus but, rather, are again mixed
with low-redshift quasars (a typical effective temperature for a white
dwarf  in the middle of  the  low-redshift quasar locus is $12000$~K).
This   suggests that the   discrimination between degenerate stars and
quasars  is bound to remain  poor.  Beyond $z~=~3.0$, the 
model~A quasars seem to remain out of the stellar locus,
and the model~C quasars stay bluer in $b-v$. This again  depends on the
particular behaviour of the Ly$\alpha$ absorbers in  the line of sight
of the observed quasar.

\begin{figure}[htb]
\centerline{\resizebox{5.4cm}{!}{\rotatebox{270}{\includegraphics{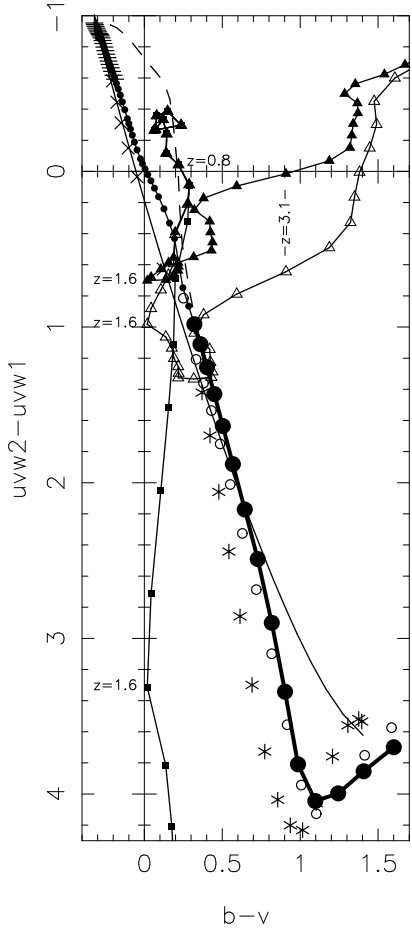}}}}
\caption{The ($uvw2-uvw1$) vs ($b-v$) colour diagram. 
The symbols have the same meaning as those used for Fig.~7. 
The highest plotted redshifts are 4.1, 3.9 and 1.8 for the 
quasars of models A, B and C respectively.}
\end{figure}

\subsection{Adding $uvw2$}

Filter   $uvw2$   could also be  used   to   build-up colour diagrams.
However,  it  should be  kept in  mind  that  the XMM-OM  is  not very
sensitive in  this passband and  the  precision of the measurement  in
$uvw2$ could be markedly worse than in any of  the other filters.  The
use of  filter  $uvw2$  is   illustrated in   Fig.~10  where  the   2D
($uvw2-uvw1$) vs ($b-v$) colour diagram   is given.  Similarly to  the
previous case, low-redshift  ($z~\leq~0.6$)  quasars are  wandering at
($uvw2-uvw1$)~=~$-0.3$  and,  of  course,  ($b-v$)~=~0.15.   This   is
slightly out  of the  stellar  locus.  However,  at  $z~\sim~0.7$, the
average  quasars progressively  become redder in  ($uvw2-uvw1$) due as
usual to  the Ly$\alpha$ emission line going  from the first filter to
the  second.  The present colour index  is expected to be discriminant
when the  Ly$\alpha$ line   is  located in  the $uvw1$  filter,  i.e.\
roughly for redshifts between 0.8 and 1.6.  This is easily seen in the
($uvw2-uvw1$) vs ($uvw1-u$)    colour   diagram as well  as   in   the
($uvw2-uvw1$) vs ($uvw1-b$) one; this pinpoints  the importance of the
joint use of  the $u$ filter  (or perhaps the $b$  one) along with the
pair   $uvw2$, $uvw1$.   Fig.~8   exhibits  the  reduced ($\sqrt{2}  /
\sqrt{4}    =  1 / \sqrt{2}$) four-dimensional     distance  in the 4D
($uvw2-uvw1$) vs ($uvw1-u$) vs ($u-b$) vs ($b-v$) colour space.  It is
clear that the contribution of the $uvw2$  filter contrasting with the
$uvw1$  one is increasing  the discrimination  power  in  the redshift
range 0.8 to 1.6.  This effect could help in generating quasar samples
that are more homogeneous in redshift since the use of the new filters
alleviates the well-known bias of $U/B$ selected quasar candidates due
to the presence of a  strong C~{\sc iv} line in  the $B$ filter (at $z
\sim 1.6 - 1.7$).  From Fig.~10, it is again clear that the degenerate
stars do not  follow the black body line  and that they still remain a
strong contaminant  of  the samples  of  quasars (particularly  around
$T_{\rm eff} \sim 13000$~K).

\subsection{General considerations}

From Fig.~8, one  can conclude that the  XMM-OM filter set is  good at
discriminating  between   non-degenerate stars   and  quasars  at  low
redshifts ($z~\leq~2.2$). This is particularly  true in the range  0.8
to   2.1 where the  use  of the  $uvw1$   and $uvw2$  filters allows a
significantly better discrimination that is even able  to wash out the
decrease in efficiency  around $z~\sim~1.6 - 1.7$ sometimes  exhibited
by  traditional  ($U-B$) vs ($B-V$)  surveys.   For very low redshifts
($z~<~0.8$), the  advantage   of  this  photometric system   is   less
marked. However, one should  not forget that  Fig.~8 gives the reduced
distance. Indeed, the minimum true distance between the quasar and the
stellar locus  is, in the 4D space  of Sect.~6.3, somewhat larger than
0.35 magnitudes (occuring   at $z=0.5$); this  already implies  a real
possibility of  segregation.  For redshifts  between  2.3 and 3.5, the
selection  is essentially inefficient,    as for ground-based  surveys
neglecting  the use of  the $R$ and  $I$ filters (for example).  It is
beneficial  to recall that  XMM-OM was originally  designed with a red
optical path that    has  been abandoned  in  the  meantime.    Beyond
$z~=~3.5$,  the  average quasar is  usually off  the stellar locus but
this corresponds  to the presence of  the Ly$\alpha$ forest in most of
the  filters  and is thus  again  highly dependent  on  the particular
realization of the Ly$\alpha$ absorption (density and actual locations
of  the   strong  Ly$\alpha$ absorbers  on   the line   of sight).  In
addition,   the flux below  Ly$\alpha$   is  comparatively much  lower
implying  a far less precise photometry.   Generally, it is clear that
the  XMM-OM filter set  is not adapted  to the  study of high-redshift
quasars: although some of them  will be easily spotted, the  selection
criterion  will  remain inhomogeneous. On   the other hand, the XMM-OM
photometry has no   discrimination power between degenerate stars  and
quasars. Particularly for white dwarfs  with effective temperatures in
the  range $12000  {\rm K}  -  15000 {\rm  K}$, the  colours are  very
similar   and  the domain  in  effective  temperature is  too small to
authorize a proper segregation.

As a last point, we  would like to  recall the existence of the $uvm2$
filter which has already been used to define the $ic_{\rm red}$ index.
We  found no combination where this  filter could  be  of some help to
improve  the   situation.  For example,   in  a   2D  ($uvw2-uvm2$)  vs
($uvm2-uvw1$) colour  diagram, the  quasars  are well located   on the
stellar   locus  except perhaps for    redshifts  around  $z~\sim  0.8
\rightarrow 1.0$ where they leave the  stellar locus but this brings no
strong improvement compared to the previously analysed filters.

\section{Conclusions}

In  the present  paper, we  discussed  the  properties of the  natural
photometric  system of  the    Optical  Monitor  onboard  the    X-ray
Multi-Mirror satellite.  On  the  basis of numerical  simulations,  we
investigated  its  transformability to   the standard Johnson  $U\!BV$
system. We gave the  main transformation equations both for individual
filters and for colour indices.  On the basis of the same technique of
simulations, we showed  that,  for stars with   effective temperatures
higher  than $9000$~K observed with   this system,  it is possible  to
determine the   temperature and  the reddening (due    to interstellar
extinction) independently, as well as to detect non-standard reddening
laws on given  lines of sight.  Finally,  we made a detailed study  of
the  possibilities to select quasar candidates  on the  basis of their
location in the   multicolour  space definable from  this   particular
photometric system.  In particular, the interest of the  use of the UV
filters has been critically  evaluated.   The main conclusion of  this
analysis is that  the use of the  $uvw1$ and $uvw2$ filters  allows to
greatly  enhance the non-degenerate  stars - quasars discrimination in
the  $[0.8;2.1]$   redshift   range.     For  lower    redshifts,  the
discrimination remains reasonably efficient. The advantage to make use
of the  XMM-OM photometric system is nevertheless  less clear for what
concerns the higher redshifts   ($z \ge 2.3$) and the   discrimination
between the quasars  and  the degenerate stars.  The  quasar candidate
selection could  benefit from the addition  of criterions based on the
X/optical flux ratio, as these are usually larger for QSO/AGN than for
stars (Stocke et al. 1991, Schmidt et al. 1998). The full treatment of
the quasar candidate selection based on XMM observations is beyond the
scope of the present paper, dedicated  to the properties of the XMM-OM
photometric system, and will be discussed elsewhere.

\acknowledgements{The authors are greatly indebted to D. Koester for the unpublished degenerate star models he kindly provided and to W. Zheng and P. Francis for a computer readable version of their composite quasar spectra. The authors also want to express their thanks to the Fonds National de la Recherche Scientifique (Belgium) 
for multiple supports. This research is supported in part by contract ARC\,94/99-178 ``Action de recherche concert\'ee de la Communaut\'e Fran\c{c}aise'' (Belgium). Partial support through the PRODEX XMM-OM Project is also gratefully acknowledged.}

\end{document}